\documentclass[preprint,12pt]{elsarticle}
\usepackage{graphics}
\usepackage{graphicx}
\usepackage{subfigure}

\usepackage{latexsym}
\usepackage{amsmath}
\usepackage{graphicx}
\usepackage{times}
\usepackage{graphicx,multicol,enumerate,subfigure,amsmath}

\usepackage{fullpage}
\usepackage{setspace}
\usepackage[english]{babel}
\usepackage[version=3]{mhchem}
\usepackage{amsmath, amssymb, amsthm}
\usepackage{verbatim, latexsym}
\usepackage{colortbl}

\newtheorem{theorem}{Theorem}
\newtheorem{remark}[theorem]{Remark}

\usepackage{soul}
\setstcolor{red}

\biboptions{sort&compress}
\journal{Journal of Computational Physics}

\onehalfspace

\begin{document}

\begin{frontmatter}

\title{ Mode Decomposition Methods for Flows in High-Contrast Porous Media. Part II. Local-Global Approach}
\author{\textbf{Mehdi Ghommem}$^1$ \corref{cor1}}
\cortext[cor1]{Email address : mehdi.ghommem@kaust.edu.sa}

\author{\textbf{Michael Presho$^{2}$, Victor M. Calo}$^{1}$, and \textbf{Yalchin Efendiev}$^{1,2}$ }

\address{$^{1}$ Applied Mathematics and Computational Science and Earth Sciences and Engineering \\
Center for Numerical Porous Media (NumPor) \\
King Abdullah University of Science and Technology (KAUST) \\
Thuwal 23955-6900, Kingdom of Saudi Arabia}

\address{$^{2}$ Department of Mathematics \& Institute for Scientific Computation (ISC) \\
Texas A\&M University \\
College Station, Texas, USA}

\begin{abstract}
In this paper, we combine concepts of the generalized multiscale finite element method and mode decomposition methods to construct a robust local-global approach
for model reduction of flows in high-contrast porous media.
This is achieved by implementing proper orthogonal decomposition (POD) and dynamic mode decomposition (DMD) techniques on a coarse grid. The resulting reduced-order approach enables a significant reduction in the flow problem size while accurately capturing the behavior of fully resolved solutions. We consider a variety of high-contrast coefficients and present the corresponding numerical results to illustrate the effectiveness of the proposed technique.
This paper is a continuation of the first part \cite{gce12} where we examine the applicability of POD and DMD to derive simplified and reliable representations of flows in high-contrast porous media. In the current paper, we discuss how these global model reduction approaches can be combined with local techniques to speed-up the simulations. The speed-up is due to inexpensive, 
while sufficiently accurate, computations of
global snapshots.

\end{abstract}


\end{frontmatter}

\section{Introduction}
Many relevant engineering and scientific applications in porous media problems consist of coupled processes in highly
heterogeneous media. For instance, media permeability can vary over several orders of magnitude due to high-conductivity fractures and/or low-conductivity shale layers. Because of these variations, the resulting large number of degrees of freedom and associated computational costs could limit the capability to perform a sensitivity analysis or conduct uncertainty quantification studies which require solving the forward problem many times. This presents the
need to develop simplified models that significantly reduce the number of degrees of freedom by
neglecting irrelevant details of the involved physics in order to remain computationally tractable.

{\bf Local multiscale methods.} Multiscale solution techniques represent a class of methods that capture the effects of small scales on a coarse grid \cite{eh09,hw97,apwy07,akl06,Calo2011,Bazilevs2007}. In this paper we follow the framework of the multiscale finite element method (MsFEM), where precomputed multiscale basis functions are used to span a coarse-grid solution space. As fine scale is embedded into the basis functions, we can recover relevant fine scale information from the multiscale solution representation. In recent years, MsFEM has been extended to allow for the systematic enrichment of coarse solution spaces in order to converge to the fine-grid solution (see e.g., \cite{EGG_MultiscaleMOR,egw11,ge10,ge10part2}). In addition, the enriched coarse spaces have been shown to be effective preconditioners in two-level domain decomposition iterative procedures \cite{ge10,ge10part2}. The flexibility associated with the construction of enriched coarse spaces makes the method an ideal solution technique for a wide variety of applications. In particular, the coarse space sizes and associated errors may be carefully calibrated in order to achieve appropriate levels of solution accuracy and computational efficiency. In situations where a higher level of accuracy is desired, additional basis functions may be incorporated in the construction of the coarse space. However, when computational efficiency is a main consideration, less basis functions may be used for a significant reduction in the coarse space dimension. In this paper we apply the enriched coarse space construction from the generalized multiscale finite element method (GMsFEM) as an effective tool for local model reduction \cite{EGG_MultiscaleMOR,egw11,ge10,ge10part2}.

{\bf Global multiscale methods.}  Two common techniques have been widely used for global model reduction, namely
dynamic mode decomposition (DMD) and proper orthogonal decomposition (POD). Both of them
are based on processing information from a sequence of snapshots (or instantaneous solutions)
to identify a low-dimensional set of basis functions. These functions are then used to derive a
low-dimensional dynamical system that is typically obtained by Galerkin projection~\cite{Akhtar2009B,Wang2011A,Wang2012A,Akhtar2012A,GhommemPCFD2012}. Proper orthogonal decomposition constitutes a powerful mode decomposition technique for extracting the most energetic structures from a linear or nonlinear dynamical process~\cite{Lumley1967,Sirovich1987A,Deane1991,Berkooz1993, Holmes1996,Akhtar2009B,Wang2011A,Wang2012A,Akhtar2012A,Hay2010A,Hay2011}. Dynamic mode decomposition has been recently proposed by Schmid \cite{Schmid2010}. In comparison with POD, this technique is intended to accurately extract the coherent and dynamically relevant structures rather than selecting the dominant modes that capture most of the flow's energy. DMD enables the computation, from simulation and empirical data, of the eigenvalues and eigenvectors of a linear model that best represents the underlying dynamics, even if those dynamics are produced by a nonlinear process. One important feature of this method is its ability to extract dynamic information from flow fields without depending on the availability of a model, but rather is based on a sequence of snapshots. As such, this technique has been successfully applied to the analysis of experimental \cite{Duke2012,Schmid2011B,Pan2011,Schmid2009,Schmid2009B,Lusseyran2011} and numerical \cite{Muld2012,Schmid2011,Schmid2010,Seena2011,Mizuno2011} flow field data and has shown a great capability to capture the relevant associated dynamics and help in the characterization of relevant physical mechanisms.

{\bf This paper.} In Part I \cite{gce12}, we followed a global approach to derive reduced-order models for flows in highly-heterogeneous porous media. This is performed by applying POD and DMD on standard finite element solutions. The DMD-based approach showed a better predictive capability due to its ability to accurately extract the information relevant for long-term dynamics, in particular, the slowly-decaying eigenmodes corresponding to largest eigenvalues. Our contribution in this work is to present a local-global model reduction framework for multiscale problems. Our approach is based on applying the aforementioned mode decomposition methods to coarse-scale problems in order to achieve a significant reduction in the system size while preserving the main flow features. In particular, we combine the concepts of GMsFEM with DMD and/or POD to develop a robust local-global approach for solving high-contrast, time-dependent parabolic problems. The resulting reduced-order method is shown to accurately capture the behavior of fully resolved solutions for a variety of high-contrast coefficients. 
Because the accuracy of both local and global approaches depends on the number
of local and global modes, when selecting local modes, our motivation
is to obtain inexpensive global snapshots while preserving the accuracy 
of the global approach. In the paper, we achieve this based on a limited
number of runs. More extensive aposteriori error estimates can guide a correct
choice for local and global modes. 
We use a variety of numerical results to illustrate the effectiveness of the proposed technique.

The remainder of the paper is organized as follows. In Sect.~\ref{modproblem} we describe the problem setting, after which we describe the local multiscale approach for model reduction in Sect.~\ref{localmodred}. Global model reduction techniques are then introduced and described in Sect.~\ref{locglobmodred}. In the same section, we also present the main steps of the local-global approach that is based on implementing mode decomposition methods on the coarse-scale problem. In Sect.~\ref{numerical} we present a variety of numerical results to complement the proposed method along with some concluding remarks in the final section.

\section{Model Problem}
\label{modproblem}
In this paper we consider a time-dependent, single-phase porous media flow governed by the following parabolic partial differential equation

\begin{equation} \label{parabolic}
\frac{\partial u}{\partial t} - \nabla \cdot \left( \kappa(x) \nabla u \right) = f(x)  \quad \text{in} \quad \Omega,
\end{equation}
where $u$ denotes the pressure, $\Omega$ is a bounded domain, $f$ is a forcing term, and $\kappa(x)$ is a positive-definite scalar function. The coefficient $\kappa$ represents the ratio of the permeability over the fluid viscosity  and is considered to be a highly-heterogeneous field with high contrast (i.e., there are large variations in the permeability). The structure of $\kappa$ is an important factor within this paper, and numerous examples will be offered in subsequent sections. We note that \eqref{parabolic} will be solved along with specified boundary and initial conditions.

\section{Local Multiscale Model Reduction}
\label{localmodred}
In this section we describe a systematic coarse grid solution technique that may be used as a reduced-order alternative to a standard fine grid approach (such as finite element discretization). The solution procedure is built within the framework of the Generalized Multiscale Finite Element Method (GMsFEM), where the solution is sought in a space of precomputed multiscale basis functions. A notable distinction of GMsFEM is that the associated coarse space may be systematically enriched to achieve a desired level of numerical accuracy  \cite{egw11,EGG_MultiscaleMOR,ge10}. In particular, the dimension of the coarse solution space may be reliably chosen based on the nature of the problem (e.g., the structure of $\kappa(x)$). The predictable accuracy of the method, combined with the inherent gain in efficiency, make GMsFEM a tractable approach for solving the model problem robustly.

\subsection{Fine grid approach}
In order to outline the fine grid solution technique, we first introduce a coarse discretization
$\mathcal{T}^H$ and assume that $\mathcal{T}^h$ is a refinement of $\mathcal{T}^H$. To solve \eqref{parabolic} using the finite element method (FEM) we search for $u_h(t) \in V^h = \text{span}\{ \phi_i \}_{i=1}^{N_f}$, where $\phi_i$ are the standard bilinear finite element basis functions defined on $\mathcal{T}^h$, and $N_f$ denotes the number of nodes on the fine grid. After multiplying the equations by test functions and integrating over the domain $\Omega$, we obtain the following set of ordinary differential equations corresponding to the model equation given in \eqref{parabolic}:

\begin{equation} \label{fem}
\mathbf{M} \frac{d\mathbf{U}}{dt} + \mathbf{A} \mathbf{U} = \mathbf{F},
\end{equation}
where $\mathbf{U} = [u_i(t)]$ denotes the time-dependent nodal solution values, $\mathbf{M}$ is the mass matrix given by $\displaystyle \mathbf{M} = [m_{ij}] = \int_\Omega \phi_i \phi_j$, $\mathbf{A}$ is the stiffness matrix given by $\displaystyle \mathbf{A} = [a_{ij}] = \int_\Omega \kappa \nabla \phi_i \cdot \nabla \phi_j$, and $\mathbf{F}$ is the forcing vector given by $\displaystyle \mathbf{F} = [f_i] = \int_\Omega f \phi_i$. Using the backward Euler, implicit scheme for the time marching process yields
\begin{equation} \label{femtime}
\mathbf{U}^{n+1} = \big( \mathbf{M} + \Delta t \mathbf{A}  \big)^{-1} \mathbf{M} \, \mathbf{U}^n +
\left( \mathbf{M} + \Delta t \mathbf{A}  \right)^{-1} \Delta t \, \mathbf{F},
\end{equation}
where $n$ denotes the time stepping index, and $\Delta t$ is the time step. The fine discretization yields large matrices of size $N_f {\times} N_f$ which may become prohibitively expensive to numerically handle.

\begin{remark}
Part I \cite{gce12} combines the standard FEM solution technique in \eqref{femtime} with methods for global model reduction. However, the focal point of the present work is to systematically reduce the dimension of the system that results from the fully-resolved (fine grid) solution. As this stage of dimension reduction hinges on the localized construction of basis functions that span the coarse solution space, we refer to it as a local approach for model reduction. By combining methods for local and global model reduction we offer a robust framework that efficiently and accurately captures the dominant features of the dynamical system for long-term forecasting.
\end{remark}

\subsection{Local multiscale approach}
\begin{figure}[htb]
  \begin{center}
      \hspace*{2.1cm} \includegraphics[width=0.62\textwidth]{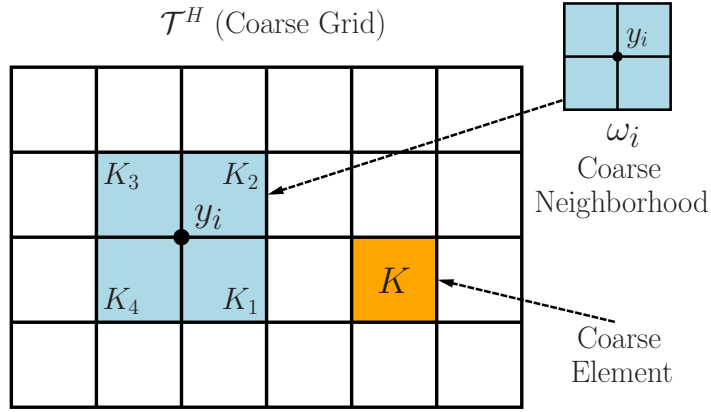}
  \end{center}
  \caption{Illustration of a coarse neighborhood}
  \label{schematic}
\end{figure}

We use $\{y_i\}_{i=1}^{N_v}$ (where $N_v << N_f$) to denote the vertices of the coarse mesh
$\mathcal{T}^H$ and define the neighborhood $\omega_i$ of the
node $y_i$ by 
\begin{equation} \label{omega}
\omega_i=\bigcup\{ K_j\in\mathcal{T}^H; ~~~ y_i\in \overline{K}_j\}.
\end{equation}
See Fig.~\ref{schematic} for an illustration of a coarse neighborhood $\omega_i$. Using the coarse mesh $\mathcal{T}^H$ we start with an initial coarse space $V_0^{\text{initial}} = \text{span}\{ \chi_i \}_{i=1}^{N_v}$, where the $\chi_i$ are standard multiscale finite element partition of unity functions satisfying
\begin{eqnarray}
-\nabla \cdot \left( \kappa(x) \nabla \chi_i \right) = 0   \quad K \in \omega_i \\
\chi_i  = g_i  \quad \text{on} \, \,  \partial K \nonumber,
\end{eqnarray}
For simplicity, for all $K \in \omega_i$, and $g_i$ is assumed to be linear.

A solution computed within $V_0^{\text{initial}}$ is the standard multiscale finite element (MsFEM) solution. However, we may systematically enrich the initial coarse space using generalized approaches for multiscale model reduction \cite{egw11,ge10,ge10part2}. In particular, we may multiply solutions resulting from localized eigenvalue problems to the initial partition of unity to enrich the coarse space. The construction of the enriched space yields that the solution error decreases with respect to the localized eigenvalue behavior. We refer the interested reader to \cite{egw11} for rigorous error estimates.

To enrich the initial coarse space, we construct the pointwise energy of the original basis functions by setting
\begin{equation}
\widetilde{\kappa} = \kappa \sum_{i=1}^{N_v} H^2 | \nabla \chi_i |^2,
\end{equation}
where $H$ denotes the coarse mesh size. Once $\widetilde{\kappa}$ is available, we solve homogeneous Neumann eigenvalue problems of the form
\begin{equation} \label{eigproblem}
-\nabla \cdot (\kappa \nabla \psi_l) = \lambda_l \widetilde{\kappa} \psi_l,
\end{equation}
on \emph{each} coarse block neighborhood $\omega_i$. We denote the eigenvalues and eigenvectors of \eqref{eigproblem} by $\{ \lambda_l^{\omega_i}  \}$ and $\{ \psi_l^{\omega_i} \}$, respectively. Since we consider a zero Neumann problem, we note that the first eigenpair is $\lambda_1^{\omega_i} = 0$ and $\psi_1^{\omega_i} = 1$. We order the resulting eigenvalues as
\begin{equation}
\lambda_1^{\omega_i} \leq \lambda_2^{\omega_i} \leq \ldots \leq \lambda_2^{\omega_i} \leq \ldots
\end{equation}
The size of the eigenvalues is closely related to the structure of $\widetilde{\kappa}$ and,
in particular, $m$ inclusions and channels in $\widetilde{\kappa}$ yields $m$ asymptotically vanishing eigenvalues. It is precisely the eigenvectors corresponding to small, asymptotically vanishing eigenvalues that we wish to use for the construction of the coarse space $V_0$. As such, we define the basis functions
\begin{equation} \label{enrichbasis}
\Phi_{i,l} = \chi_i \psi_l^{\omega_i} \quad \text{for} \, \, 1 \leq i \leq N_v ~~\text{and} ~~1 \leq l \leq L_i,
\end{equation}
\noindent
where $L_i$ denotes the number of eigenvectors that will be chosen for each node $i$. With the updated basis functions in place, we define the local spectral multiscale space as

\begin{equation}
V_0 = \text{span}\{ \Phi_{i,l}:~~1 \leq i \leq N_v ~~\text{and} ~~1 \leq l \leq L_i \}.
\end{equation}
Using a slightly different index notation, we may write $V_0 = \text{span}\{ \Phi_{i} \}_{i=1}^{N_c}$, where $N_c$ denotes the total number of basis functions used in the coarse space construction.

Through constructing an operator matrix $\mathbf{R}_0^T = [ \Phi_1, \ldots, \Phi_{N_c} ]$ (where $\Phi_i$ are taken to be the nodal values of each basis function defined on the fine grid), we can express the coarse scale analogue of \eqref{femtime} as
\begin{equation} \label{msfemtime}
\mathbf{U}_0^{n+1} = \big( \mathbf{M}_0 + \Delta t \mathbf{A}_0  \big)^{-1} \mathbf{M}_0 \, \mathbf{U}_0^n +
\left( \mathbf{M}_0 + \Delta t \mathbf{A}_0  \right)^{-1} \Delta t \, \mathbf{F}_0,
\end{equation}
where $\mathbf{M}_0 = \mathbf{R}_0 \mathbf{M}  \mathbf{R}_0^T$, $\mathbf{A}_0 = \mathbf{R}_0 \mathbf{A}  \mathbf{R}_0^T$, and $\mathbf{F}_0 = \mathbf{R}_0 \mathbf{F}$. To elaborate on the form in \eqref{msfemtime}, we reiterate that we now seek solutions within the spectral multiscale space $V_0$. A more detailed consideration of the resulting coarse scale matrices (using the mass matrix as a specific example) yields an expression of the form

\begin{equation*}
\mathbf{M}_0 = [m^0_{IJ}] = \int_\Omega \Phi_I \Phi_J = \mathbf{R}_0 \mathbf{M}  \mathbf{R}_0^T,
\end{equation*}
since for all $p,q \in V^h$ we have $\displaystyle p^T \mathbf{M} q = \int_\Omega pq$. $\mathbf{R}_0$ analogously allows for the extension of coarse scale solutions onto the fine grid. The resulting coarse matrices in \eqref{msfemtime} are of size $N_c {\times} N_c$, where $N_c$ is significantly smaller than $N_f$. Thus, the coarse system in \eqref{msfemtime} offers a suitable local model reduction of the fine system in \eqref{femtime}.

\section{Global Model Reduction}
\label{locglobmodred}
Two mode decomposition methods are used in this study for global model reduction, namely Proper
Orthogonal Decomposition (POD) and Dynamic Mode Decomposition (DMD).
The basic principles of POD and DMD are presented in this section. For
more detailed description of POD, the reader is refereed to \cite{Lumley1967,Sirovich1987A,Deane1991,Berkooz1993, Holmes1996,Akhtar2009B,Wang2011A,Wang2012A,Akhtar2012A,Hay2010A,Hay2011} and for DMD \cite{Schmid2010,Schmid2011,Schmid2011B,Chen2012}.

The first step for either the POD- or DMD- analysis is to collect a sequence of $N$ instantaneous data fields (or snapshots) $\textbf{v}_j$ given by
\begin{eqnarray}
\textbf{V}_1^N=\{\textbf{v}_1,\textbf{v}_2,\textbf{v}_3,\cdots, \textbf{v}_N\}
\label{Eq1}
\end{eqnarray}
where the time spacing between two consecutive snapshots in the above sequence is assumed to be constant. The POD proceeds by performing a singular value decomposition of the sequence of snapshots $\textbf{V}_1^N$. To this end, we first form the correlation matrix $C$ from the snapshot sequence as:
\begin{eqnarray}
C=(\textbf{V}_1^N)^{T}\;\textbf{V}_1^N,
\label{Eq2}
\end{eqnarray}
and then compute the POD modes $\phi^{POD}$ by performing an eigen-analysis of the correlation matrix $C$; that is,
\begin{eqnarray}
C \;W_i = \sigma_i^2 \;W_i \quad \mbox{and} \quad \phi^{POD}_i=\frac{1}{\sigma_i}\textbf{V}_1^N\; W_i.
\label{Eq3}
\end{eqnarray}
The selection of POD modes is based on an energy ranking of the coherent structures. However, the energy may not in all circumstances be the appropriate measure to rank the importance of the flow structures and detect the most dynamically-relevant modes \cite{gce12,Chen2012}.

The DMD method is based on postprocessing the sequence of snapshots $\textbf{V}_1^N$ to extract the dynamic information. It uses the Arnoldi approach to relate two consecutive data fields through a linear mapping; that is,
\begin{eqnarray}
\textbf{v}_{i+1}=\textbf{A}\textbf{v}_i. \label{Eq4}
\end{eqnarray}
The eigenvalues and eigenvectors of the matrix $\textbf{A}$ characterize the dynamic behavior of the flow. In practical situations, the matrix $\textbf{A}$ might be very large or even its exact form may not be given. As such, the DMD method enables the approximation of the relevant eigenvalues and eigenvectors of the matrix $\textbf{A}$. To proceed with DMD, we first assume that the last snapshot can be represented by a linear combination of the previous snapshots; that is,
\begin{eqnarray}
\textbf{v}_N = \sum_{i=1}^{N-1} a_i \textbf{v}_i + \textbf{r} \label{Eq5}
\end{eqnarray}
or
\begin{eqnarray}
\textbf{v}_N = \textbf{V}_1^{N-1} \; \textbf{a} + \textbf{r} \label{Eq6}
\end{eqnarray}
where $\textbf{a}=\{a_1,a_2,\cdots,a_{N-1}\}^T$ and $\textbf{r}$ is the residual vector. This yields
\begin{eqnarray}
\textbf{A}\; \textbf{V}_1^{N-1} = \textbf{V}_2^{N} = \textbf{V}_1^{N-1} \; \textbf{S}+ \textbf{r} \; \textbf{e}^T_{N-1} \label{Eq7}
\end{eqnarray}
where $\textbf{e}^T_{N-1}=\left(
                            \begin{array}{cccc}
                              0 & \cdots & 0 & 1 \\
                            \end{array}
                          \right)
$ is the $(N-1)$ unit vector and the matrix $\textbf{S}$ is of companion type defined as:
\begin{eqnarray}
\textbf{S} =\left(
                   \begin{array}{ccccc}
                     0 &  &  &  & a_1 \\
                     1 & 0 &  &  & a_2 \\
                      & \ddots & \ddots &  & \vdots \\
                      &  & 1 & 0 & a_{N-2} \\
                      &  &  & 1 & a_{N-1} \\
                   \end{array}
                 \right). \label{Eq8}
\end{eqnarray}
The unknown matrix $\textbf{S}$ is determined by minimizing the residual $\textbf{r}$. The minimization problem expressed as
\begin{eqnarray}
\textbf{S}=\min_{\textbf{S}} \parallel \textbf{V}_2^{N} - \textbf{V}_1^{N-1}\textbf{S} \parallel \label{Eq9}
\end{eqnarray}
can be solved using a QR-decomposition of the matrix $\textbf{V}_1^{N-1}$. Once $\textbf{S}$ is determined, we compute its eigenvalues
and eigenvectors can be computed which result in the DMD modes $\phi^{DMD}$ as follows:
\begin{eqnarray}
\textbf{S}= \textbf{R}^{-1} \; \textbf{Q}^T \; \textbf{V}_2^{N} \label{Eq9}
\end{eqnarray}
\begin{eqnarray}
\phi^{DMD}_j = \textbf{V}_1^{N-1} \; X_j \label{Eq10}
\end{eqnarray}
where
\begin{eqnarray}
\textbf{S} X_j= \lambda_j X_j. \label{Eq11}
\end{eqnarray}
A more robust approach to compute the DMD modes \cite{Schmid2010} is based on a preprocessing step using a singular value decomposition of the data sequence $\textbf{V}_1^{N-1} = \textbf{U} \; \Sigma \; \textbf{W}^T $. Substituting the SVD representation $\textbf{U} \; \Sigma \; \textbf{W}^T $ into Eq. (\ref{Eq7}) and multiplying the result by from the left $\textbf{U}^T$ and by $\textbf{W} \; \Sigma^{-1}$ from  the right, we obtain the following matrix
\begin{eqnarray}
\textbf{U}^T \; \textbf{A} \; \textbf{U} = \textbf{U}^T \; \textbf{V}_2^{N} \; \textbf{W} \; \Sigma^{-1} \equiv \tilde{\textbf{S}}. \label{Eq15}
\end{eqnarray}
The dynamic modes are then computed from the matrix $\tilde{\textbf{S}}$ as follows:
\begin{eqnarray}
\phi^{DMD}_j = \textbf{U} \; \textbf{y}_j, \label{Eq16}
\end{eqnarray}
where $\textbf{y}_j$ is the $j^{\mbox{th}}$ eigenvector of $\tilde{\textbf{S}}$, i.e., $\tilde{\textbf{S}}\; \textbf{y}_j =\mu_j \textbf{y}_j$, and $\textbf{U}$ is the matrix collecting the right singular vectors of the snapshot sequence $\textbf{V}_1^{N-1}$. The latter approach is used in the present study.
\section{Numerical Results}
\label{numerical}
\begin{figure}[htb]
  \begin{center}
      \subfigure[Case I: high-conductivity channels]{\includegraphics[width=0.48\textwidth]{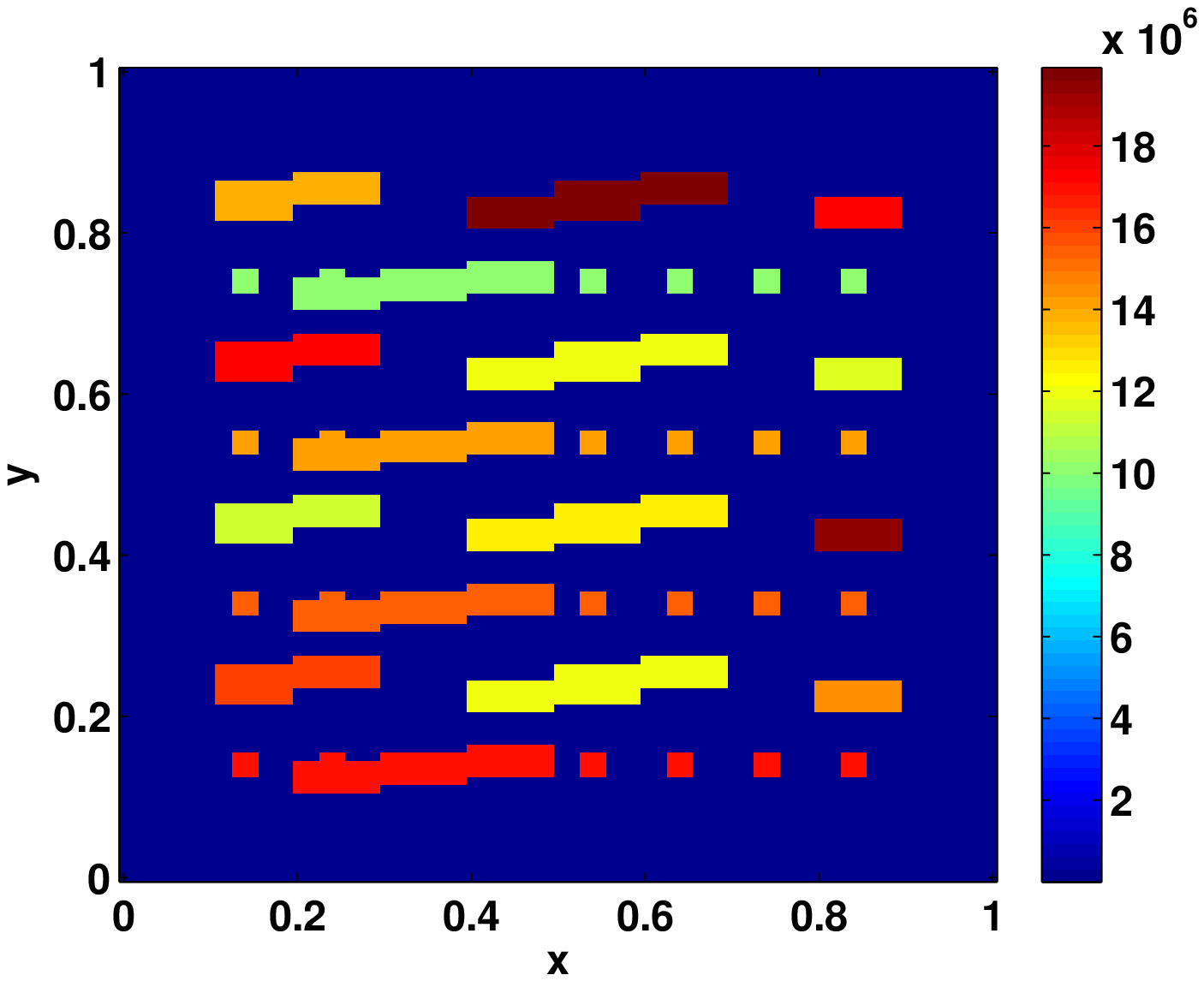}\label{Per_High}}
      \subfigure[Case II: low-conductivity layers]{\includegraphics[width=0.48\textwidth]{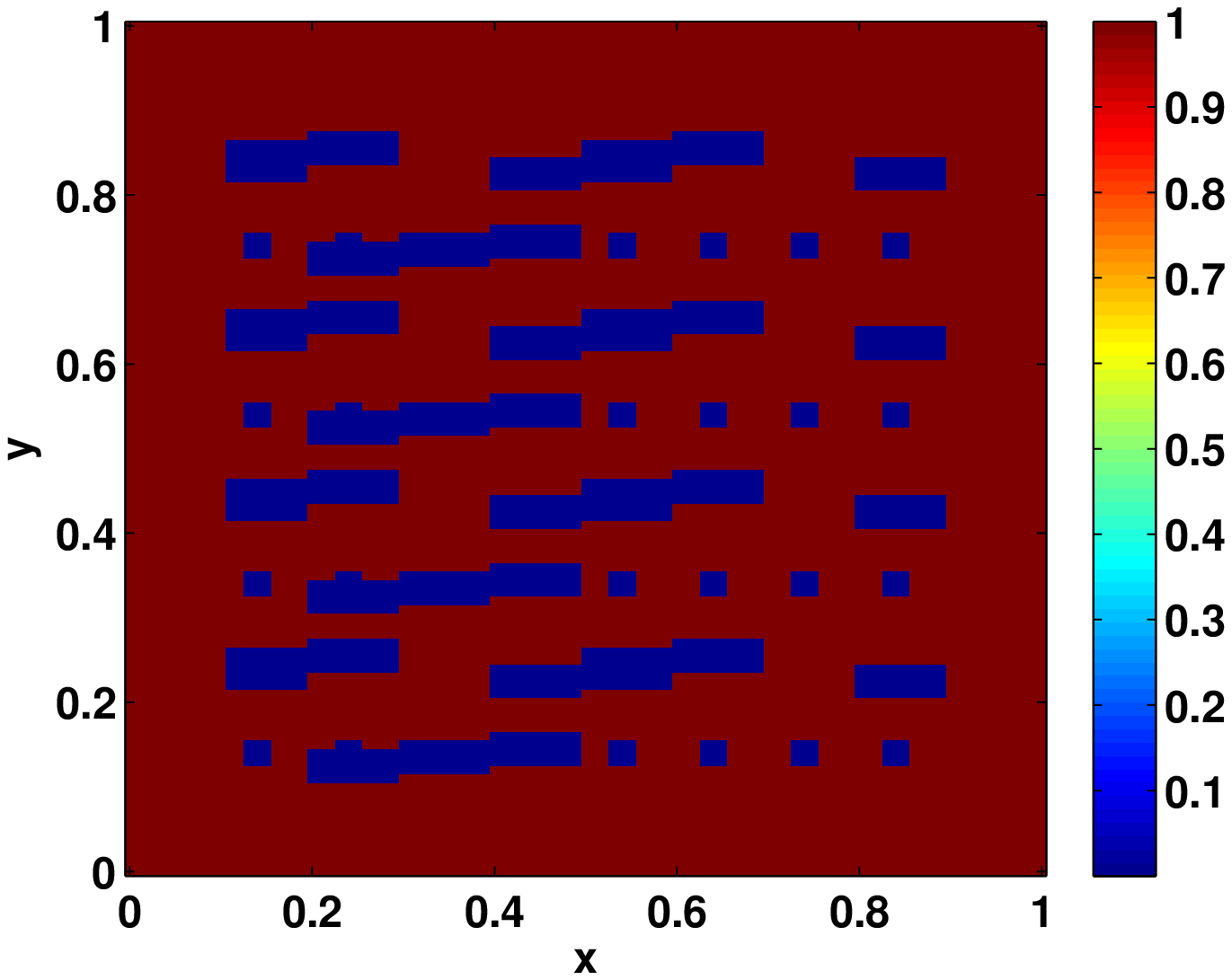}\label{Per_Low}}
  \end{center}
  \caption{Different configurations of the permeability field; high conductivity channels (left), low
                conductivity layers (right). }
  \label{Permeability}
\end{figure}

\begin{figure}[htb]
  \begin{center}
      \includegraphics[width=0.6\textwidth]{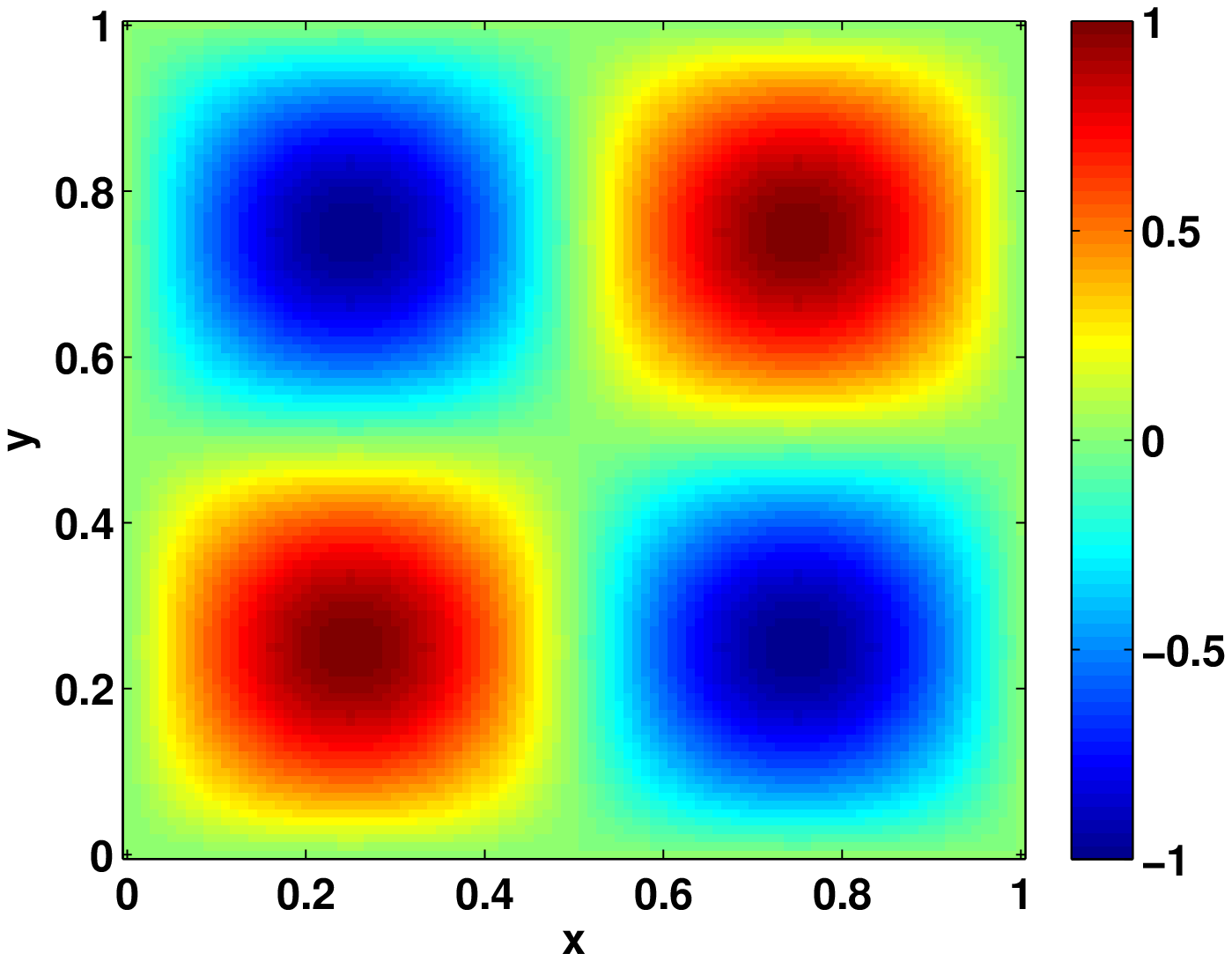}
  \end{center}
  \caption{Initial configuration of the solution field.}
  \label{Init1}
\end{figure}

In this section, we analyze the effects of permeability fields that contain inclusions of high and low conductivity as shown in Fig. \ref{Permeability}.  These configurations lead to different types of dynamical behavior. Fig. \ref{Per_High} is a field that models high conductivity channels within an otherwise homogeneous domain. The minimum conductivity value for this case is taken to be $\kappa_{\min} = 1$, and the high conductivity channels vary randomly (from channel to channel) with a maximum value of $\kappa_{\max} = 1.9{\times}10^7$. Fig. \ref{Per_Low} is a field of similar structure, yet the permeability values are inverted such that the layers (similarly placed) now represent low conductivity regions within a high conductivity value. Both permeability fields are described by a $100{\times}100$ fine mesh. Using such configurations, we solve the problem governed by Eq. (\ref{msfemtime}) on a coarse grid over a time interval of $T = [0 \; 1]$. This time interval is observed to be long enough so that the steady-state solution for all relevant cases is achieved. We use a time step of $\Delta t = 5{\times}10^{-4}$, and solve the model on a two-dimensional unit domain $\Omega = [0,1]{\times}[0,1]$. We assume zero Dirichlet boundary conditions, and the initial solution configuration is shown in Fig. \ref{Init1}.

\subsection{Local multiscale model reduction results}
\begin{figure}[htb]
\centering
\subfigure[Fine ($N_f = 10201$)]{\hspace*{-0.3cm} \includegraphics[width=0.33\textwidth]{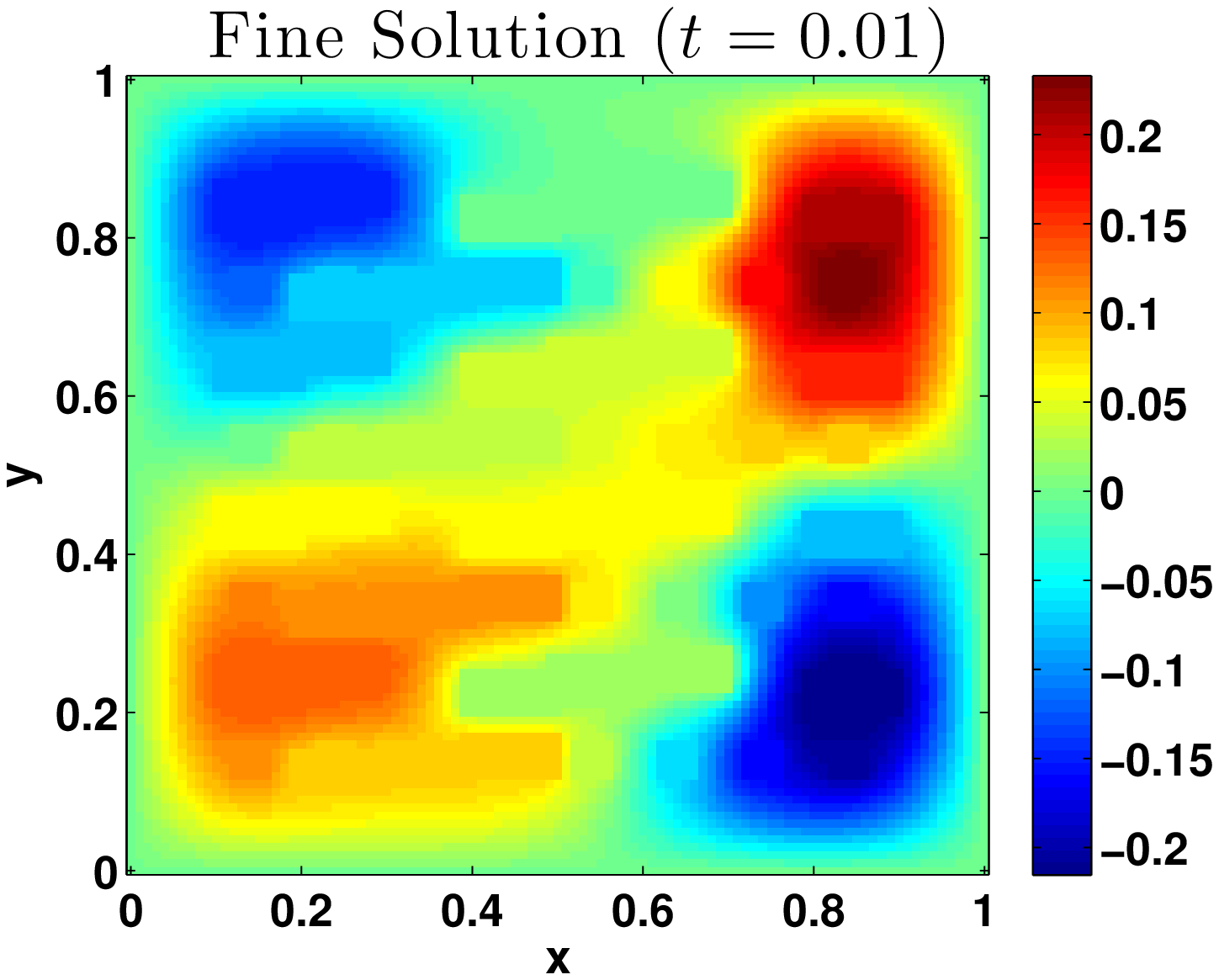}
                                                      \includegraphics[width=0.33\textwidth]{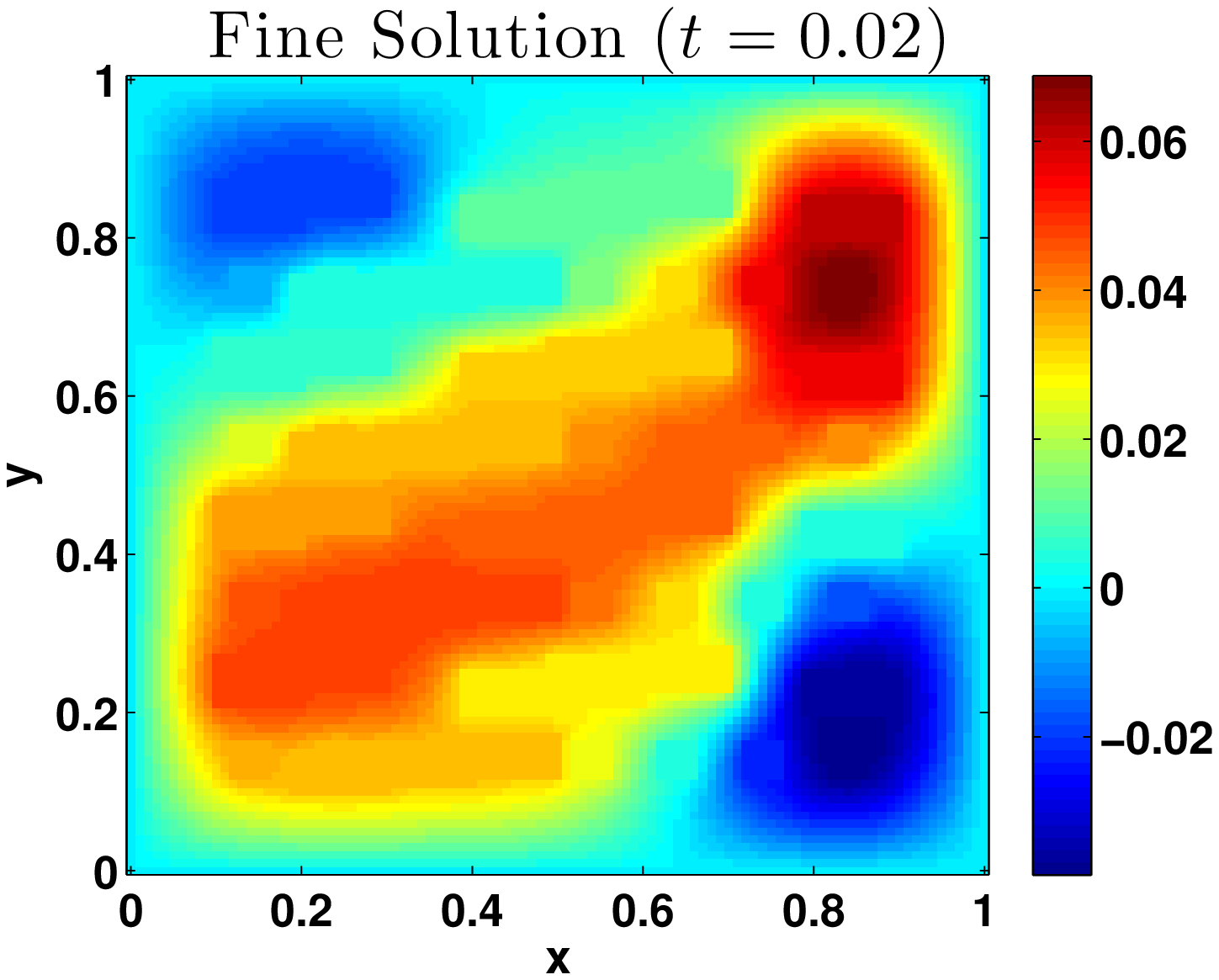}
                                                      \includegraphics[width=0.33\textwidth]{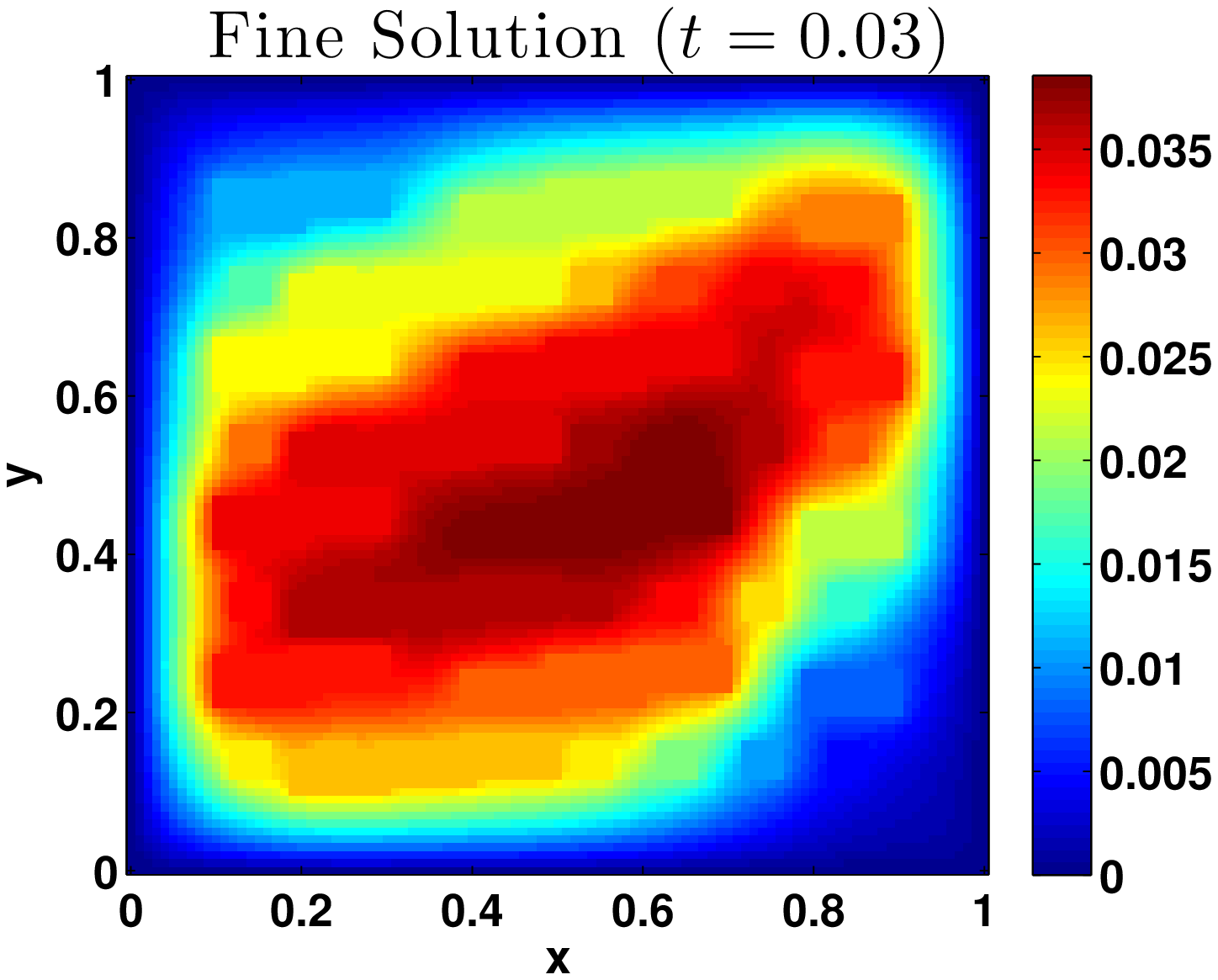} }

 \subfigure[Coarse ($N_c = 850$)]{\hspace*{-0.3cm} \includegraphics[width=0.33\textwidth]{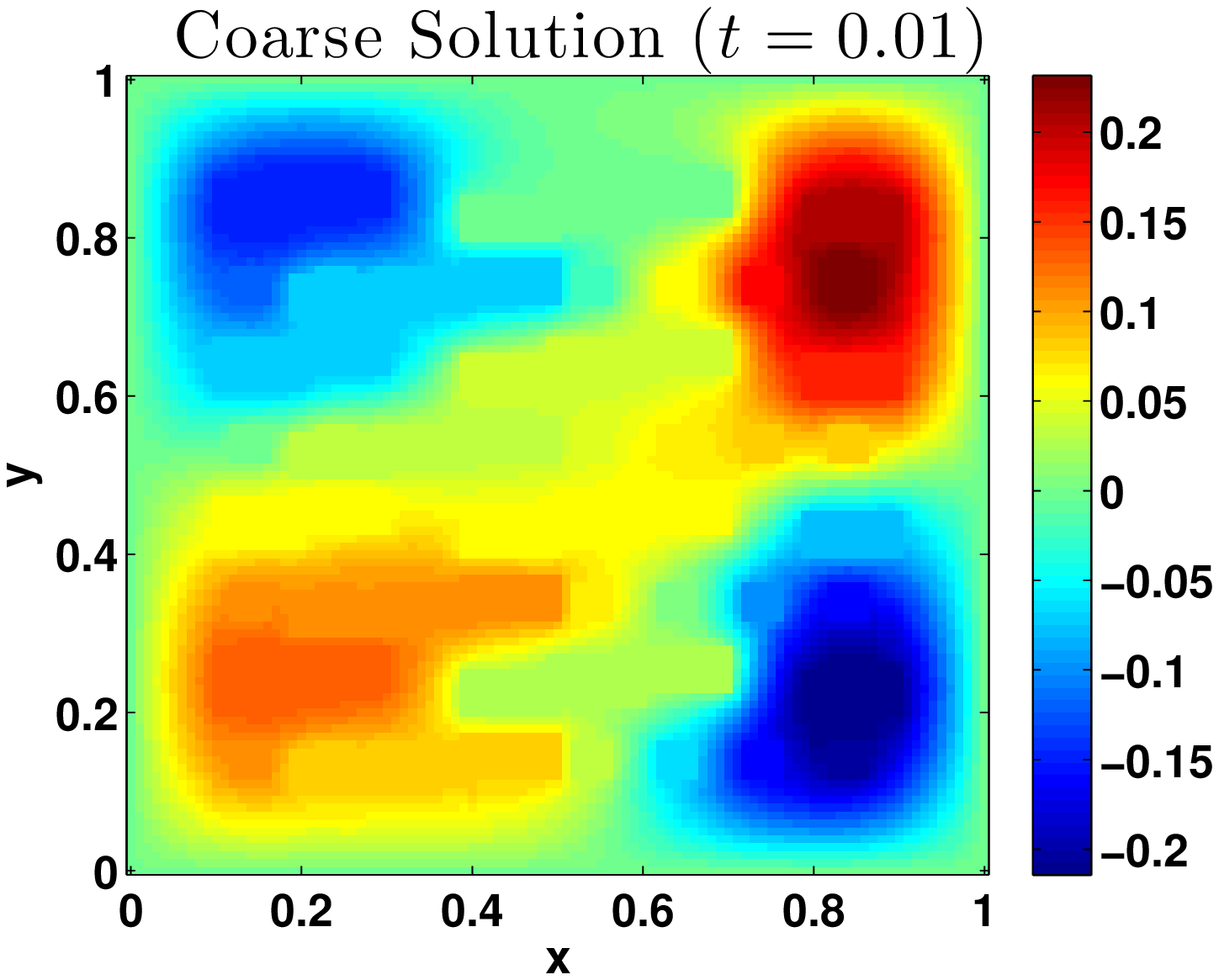}
                                                           \includegraphics[width=0.33\textwidth]{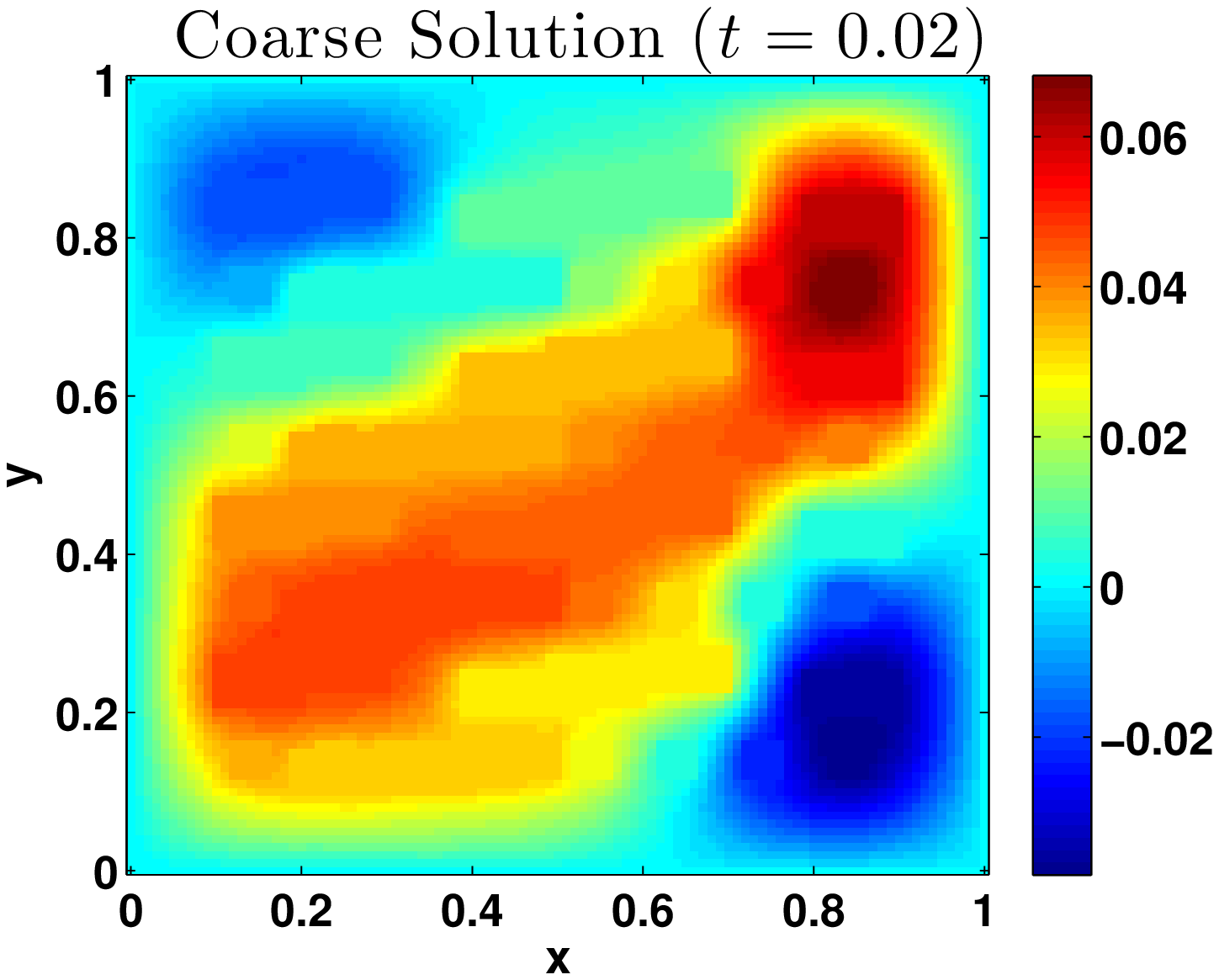}
                                                           \includegraphics[width=0.33\textwidth]{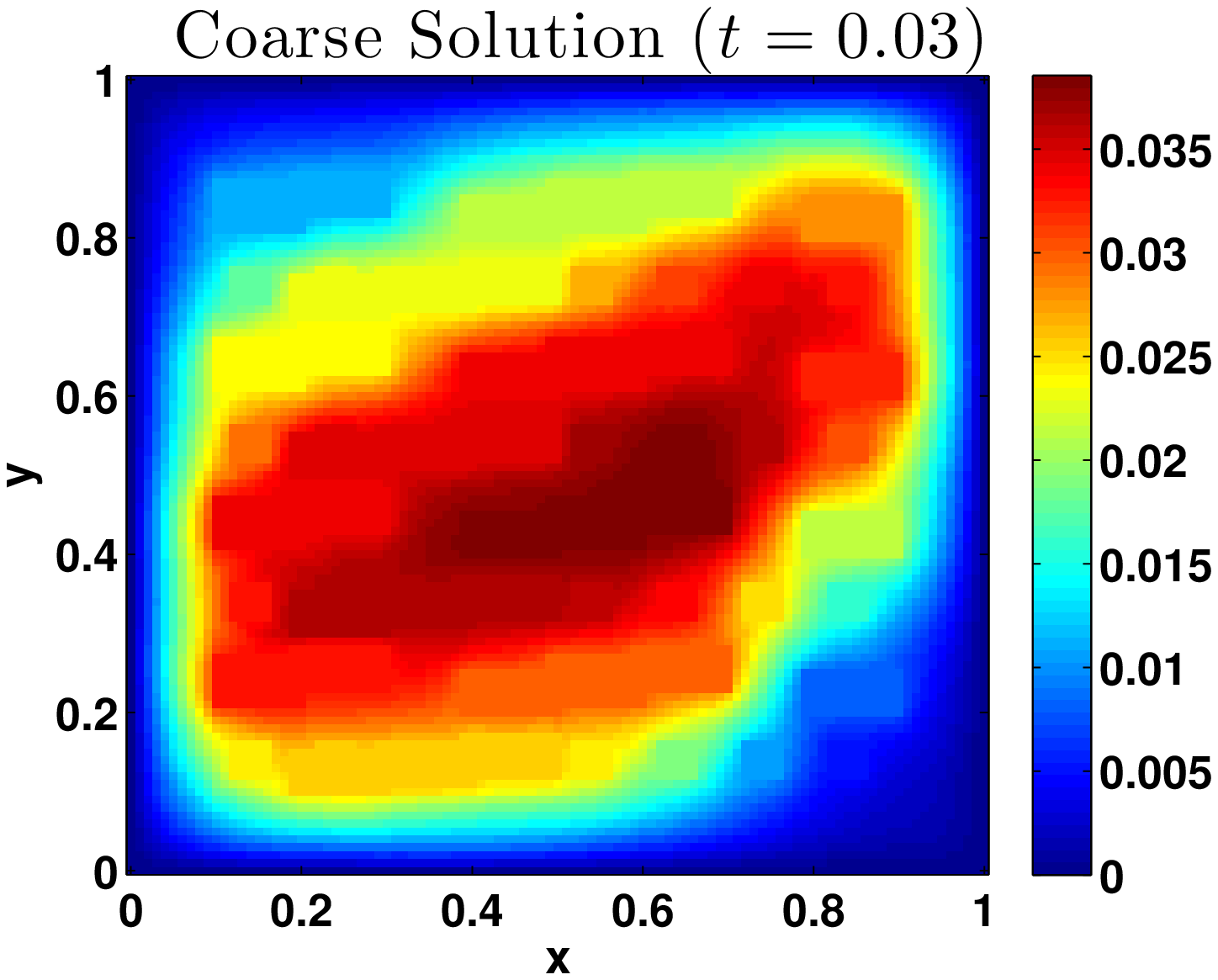} }

  \caption{Fine (top) and coarse (bottom) solutions advancing in time }
  \label{finevscoarse}
\end{figure}

\begin{figure}[htb]
   \centering
   \includegraphics[width=0.5\textwidth]{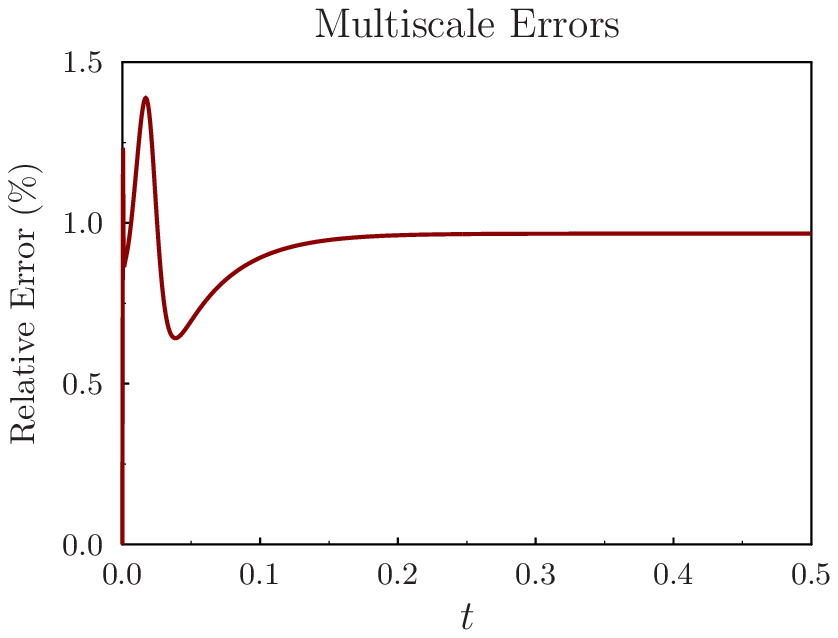}
   \caption{Relative errors between fine and multiscale solutions advancing in time }
   \label{errors}
\end{figure}

To further motivate the combined local-global solution procedure, we first present results from the local multiscale approach within this subsection. As the local-global approach described in Sect.~\ref{locglobmodred} does not require solutions that are obtained through a coarse grid approximation (see \cite{gce12}), the addition of an effective multiscale model will offer a more efficient solution technique which is robust. For these initial comparisons we use a forcing term $f=1$, the permeability field from Fig.~\ref{Per_High}, and recall that the fine solutions are obtained from Eq.~\eqref{femtime} and the multiscale (coarse) solutions are obtained from Eq.~\eqref{msfemtime}. For these examples, the fine mesh yields a system of size $N_f = 10201$, and the local multiscale approach yields a system of size $N_c = 850$. Thus, we are solving a reduced-order system which is an order of magnitude smaller than its fully-resolved counterpart. See Fig.~\ref{finevscoarse} for an illustration of fine and reduced-order solutions advancing in time. The solution profiles are nearly indistinguishable from one another, formally validating the success of the local multiscale approach. For a more rigorous comparison, we also offer the relative $L_2$ error quantities $\| u^{N_f}(t) -  u^{N_c}(t) \| ~~/~~\| u^{N_f}(t) \| \times 100 \%$ for a representative amount of time in Fig.~\ref{errors}.

The error values in Fig.~\ref{errors} typically range from $1.0-1.5 \%$ for any time value. This range of values results from a careful choice of the coarse space dimension that we use in the local model reduction. In particular, we choose the dimension of the coarse space such that the errors associated with GMsFEM are comparable to those that will result from the respective global model reduction technique. In doing so, we avoid  the larger computations that are associated with achieving a more strict level of accuracy (due to the addition of more basis functions in the coarse space construction). The alternative would be to devote unnecessary resources to obtain solutions whose errors would be essentially negligible compared to those those resulting from the global model reduction. So while it may be possible to compute more accurate solutions within a larger coarse space, such an effort would be in excess of the optimal execution of the proposed method.

\subsection{Local-global model reduction results}
 The variations of the forcing term $f$ in our numerical examples over $\Omega$ is shown in Fig. \ref{Forcing}. The objective is to investigate the capability of a combination of global model reduction techniques, POD and DMD, with local multiscale approach to capture the main flow characteristics and reproduce the flow dynamics response within a certain accuracy and at a reduced computational cost. The local-global approach involves the following steps:

 \begin{figure}[htb]
 \begin{center}
 \includegraphics[width=0.6\textwidth]{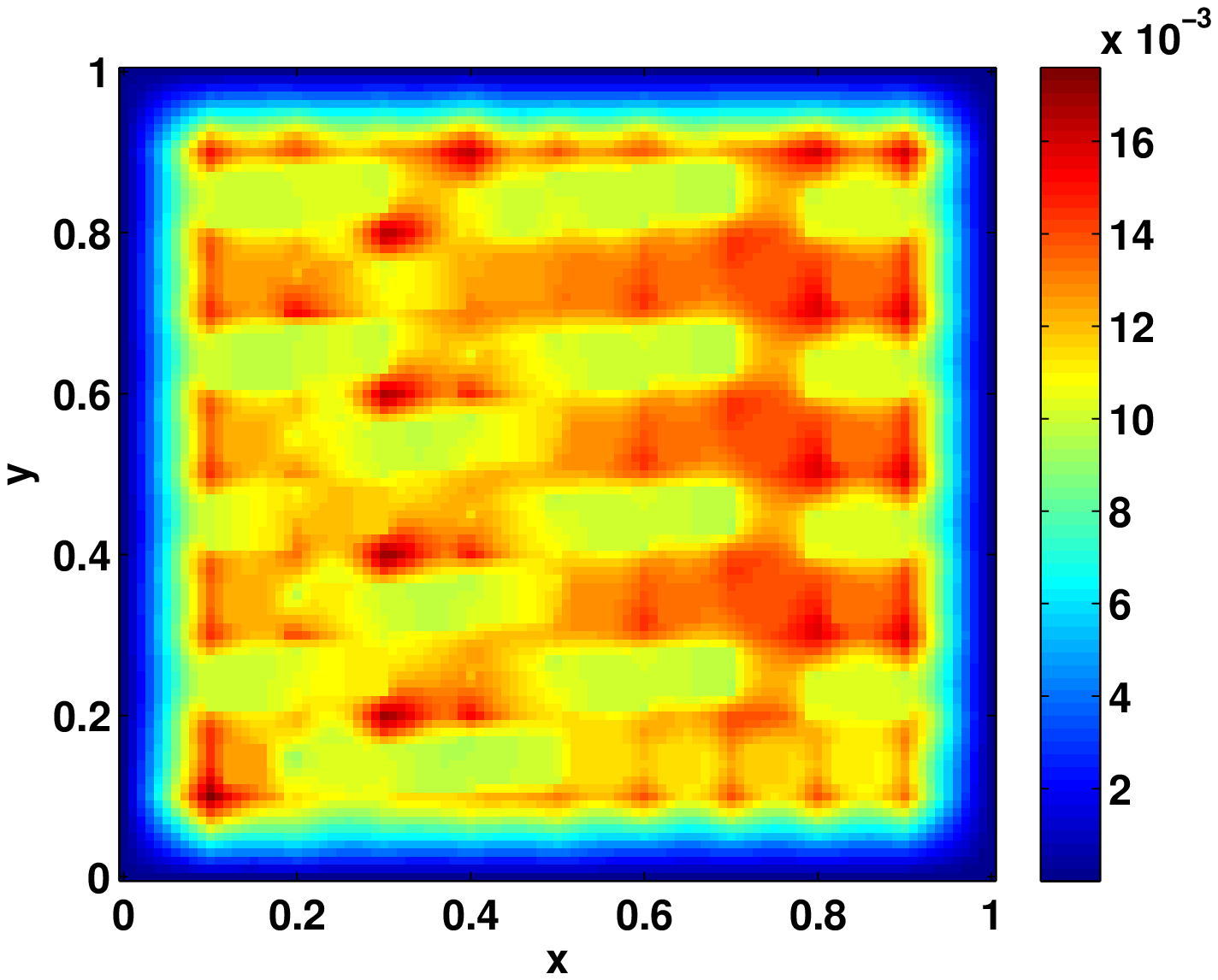}
 \end{center}
\caption{Spatial variations of the forcing $f$ over the domain $\Omega$.}\label{Forcing}
\end{figure}

\begin{itemize}
\item  Using the same solution parameters from above, we record $N_t$ instantaneous solutions (usually referred as snapshots) and collect them in a snapshot matrix as:
\begin{eqnarray}
\textbf{V}_1^{N_t}=\{\textbf{v}_1,\textbf{v}_2,\textbf{v}_3,\cdots, \textbf{v}_{N_t}\}
\end{eqnarray}
where $N_t$ is the number of snapshots and $N_c$ is the size of the column vectors $\textbf{v}_{i}$. A particular attention should be drawn when selecting the set of snapshots that will be used to extract the modes \cite{gce12}. For both cases, we start collecting snapshots from the 10$^{\mbox{th}}$ time step. The use of 25 snapshots is observed to be appropriate to reproduce results of Case I (inclusions with high conductivity) with acceptable accuracy. However, 100 snapshots were required to properly analyze Case II. This is expected since the permeability field that corresponds to Case II contains inclusions of low conductivity and subsequently involves longer time dynamics.

\item We postprocess the snapshot matrix, as described in the previous section, to compute the POD and DMD modes and use these modes to approximate the solution field in the coarse grid. As such, we assume an expansion in terms of the modes $\phi^k_i$; that is, we let
\begin{equation}
u_0(x,t) \approx \tilde{u}_0(x,t) = \sum_{i=1}^{N_r} \alpha_i(t) \phi^k_i(x)\label{utilda}
\end{equation}
or in a matrix form
\begin{equation}
\mathbf{U}_0^n \approx \tilde{\mathbf{U}}_0^n = \Phi \alpha^n \label{utildam}
\end{equation}
where $\Phi=\left(
              \begin{array}{ccc}
                \phi_1^k & \cdots & \phi_{N_r}^k \\
              \end{array}
            \right)
$ and $k$ can refer to POD or DMD. For the sake of comparison, we extract and keep the same number of modes for POD and DMD in our simulations. We consider the use of the first six modes for all simulations. In fact, the singular value decomposition of the snapshot matrix showed that the first six POD modes contain more than 99.9\% of the total flow energy.
\item To assess the capability of the POD and DMD modes in capturing the dynamics involved in the process and enabling good projection subspaces, we evaluate the $L_2$ projection error; that is, we project each snapshot onto the POD modes, compute the following inner product
\begin{equation}
\alpha_i(t_j)=\left< \phi^k_i(x),u_0(x,t_j)\right>, \;\;\mbox{or}\;\; \alpha^j=(\Phi^*\Phi)^{-1}\Phi^*\mathbf{U}_0^j  \label{Eq19}
\end{equation}
where $\left< F,G\right>=\int_{\Omega}(F\; G) \;d\Omega$, and define the relative error as the $L_2$-norm of the difference between the reference and approximate solutions over the reference one; i.e.
\begin{equation}
\parallel E(t) \parallel_2 = \frac{\parallel \tilde{u}_0(x,t) - u_0(x,t) \parallel_2}{\parallel u_0(x,t) \parallel_2}. \label{Eq20}
\end{equation}

A low projection error indicates the ability of modal decomposition techniques to compute good projection subspaces, but it does not necessarily yield stable and accurate reduced-order models when Galerkin projection is used.

\item We use the solution expansion given by Eq. (\ref{utilda}) and project the governing equation of the coarse scale problem onto the space formed by the modes to obtain a set of $N_r$ ordinary differential equations that constitute a reduced-order model; that is,
\begin{equation}\label{rom}
\dot{\alpha} = - (\Phi^*\mathbf{M}_0\Phi)^{-1} \Phi^* \mathbf{A}_0 \Phi \alpha + (\Phi^*\mathbf{M}_0\Phi)^{-1} \Phi^* \mathbf{F}_0.
\end{equation}
Thus, the original problem with $N_f$ degrees of freedom is reduced to a dynamical system with $N_r$ dimensions where $N_r \ll N_f$.
\item We use the operator matrix $\mathbf{R}_0$ to downscale the approximate solution and evaluate the flow field in the fine scale domain.
\end{itemize}

\begin{figure}[htb]
  \begin{center}
      \includegraphics[width=0.6\textwidth]{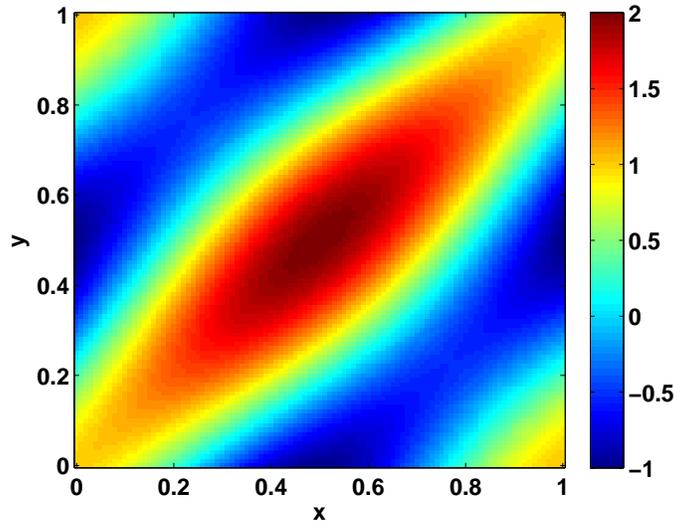}
  \end{center}
  \caption{Initial configuration of the solution field.}
  \label{Init2}
\end{figure}

\begin{figure}[htb]
  \begin{center}
      \subfigure[case I: high-conductivity channels]{\includegraphics[width=0.48\textwidth]{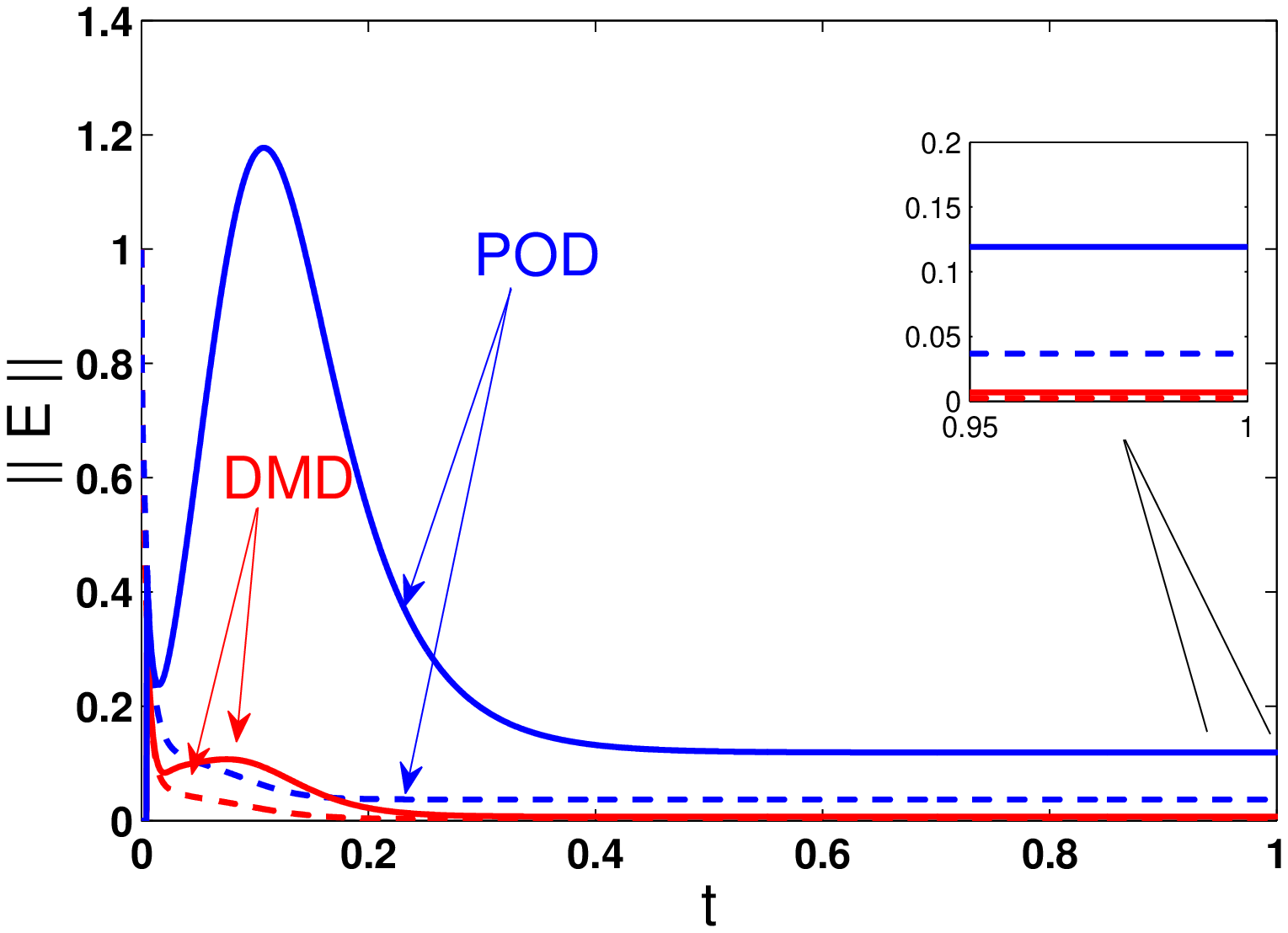}\label{Error_HighCond}}
      \subfigure[case II: low-conductivity layers]{\includegraphics[width=0.48\textwidth]{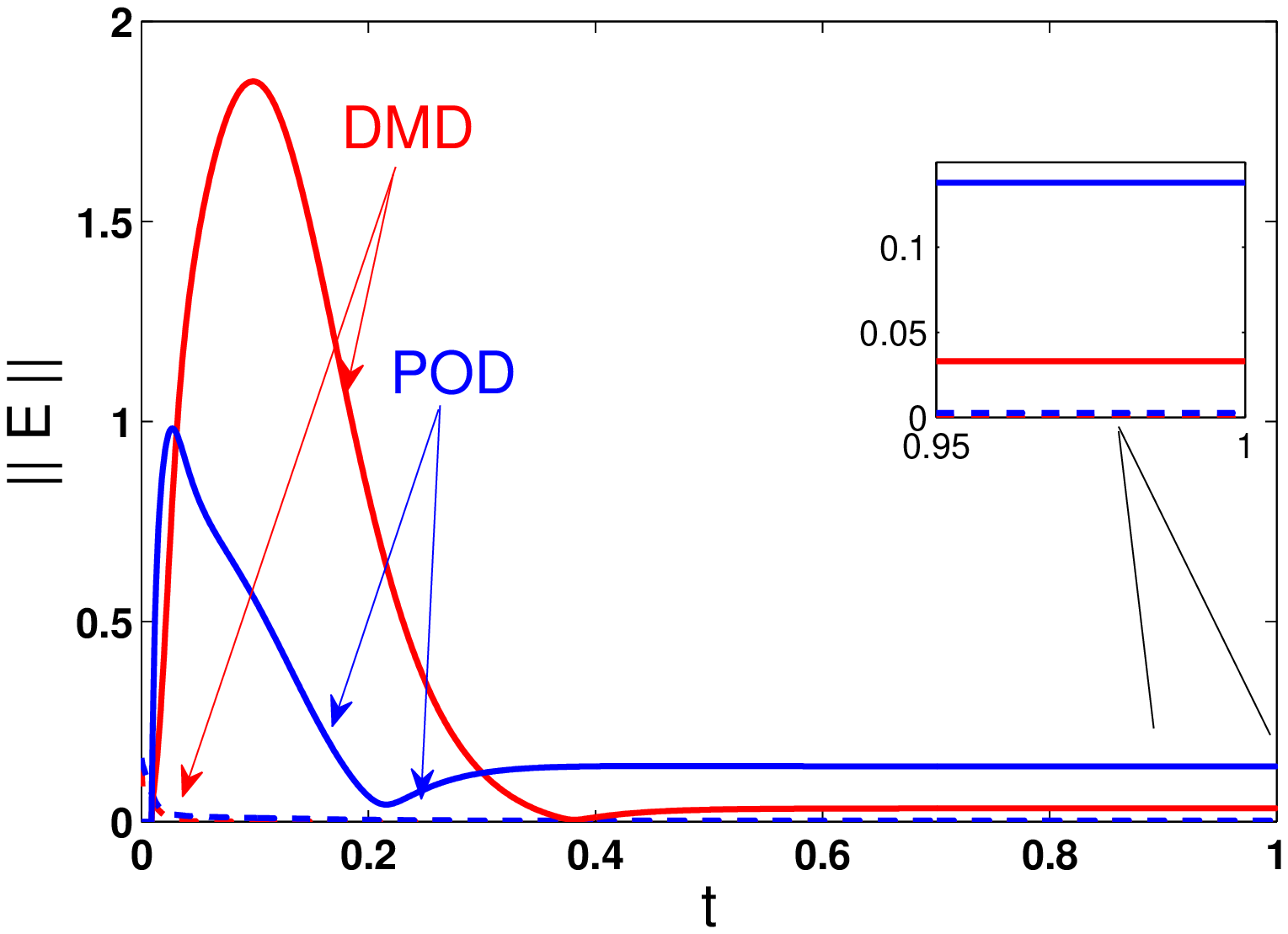}\label{Error_LowCond}}
  \end{center}
  \caption{Variations of the Galerkin projection error (solid line) and L$_2$ projection error (dashed line) with time for different permeability configurations. Results are obtained using POD and DMD modes.}
  \label{Error}
\end{figure}

We follow the above steps to derive reduced-order models while considering the different initial configuration shown in Fig. \ref{Init2}. This is performed intentionally to test the robustness of the reduced-order model with respect to variations in the initial conditions. In Fig. \ref{Error}, we plot the variations of the $L_2$ projection and Galerkin projection errors with time for Cases I and II. Results are obtained from the DMD- and POD-based approaches. We observe small $L_2$ projection errors. However, larger errors are reached when projecting the governing equations onto the subspace spanned by the modes to obtain a reduced-order model. In particular, we observe jumps to high values during the first time steps. This is due to the use of different initial configuration. For both cases of high and low conductivity, the Galerkin projection errors decrease and reach relatively small values as time evolves and the steady state develops (Case I: 12\% for POD and 0.6 \% for DMD and Case II: 14\% for POD and 3 \% for DMD). This indicates the suitability of the combined local-global approach for model reduction of flows in highly heterogeneous porous media.

\begin{figure}[htb]
  \begin{center}
      \subfigure[Fine -Reference solution ($N_f=10201$)]{\includegraphics[width=0.32\textwidth]{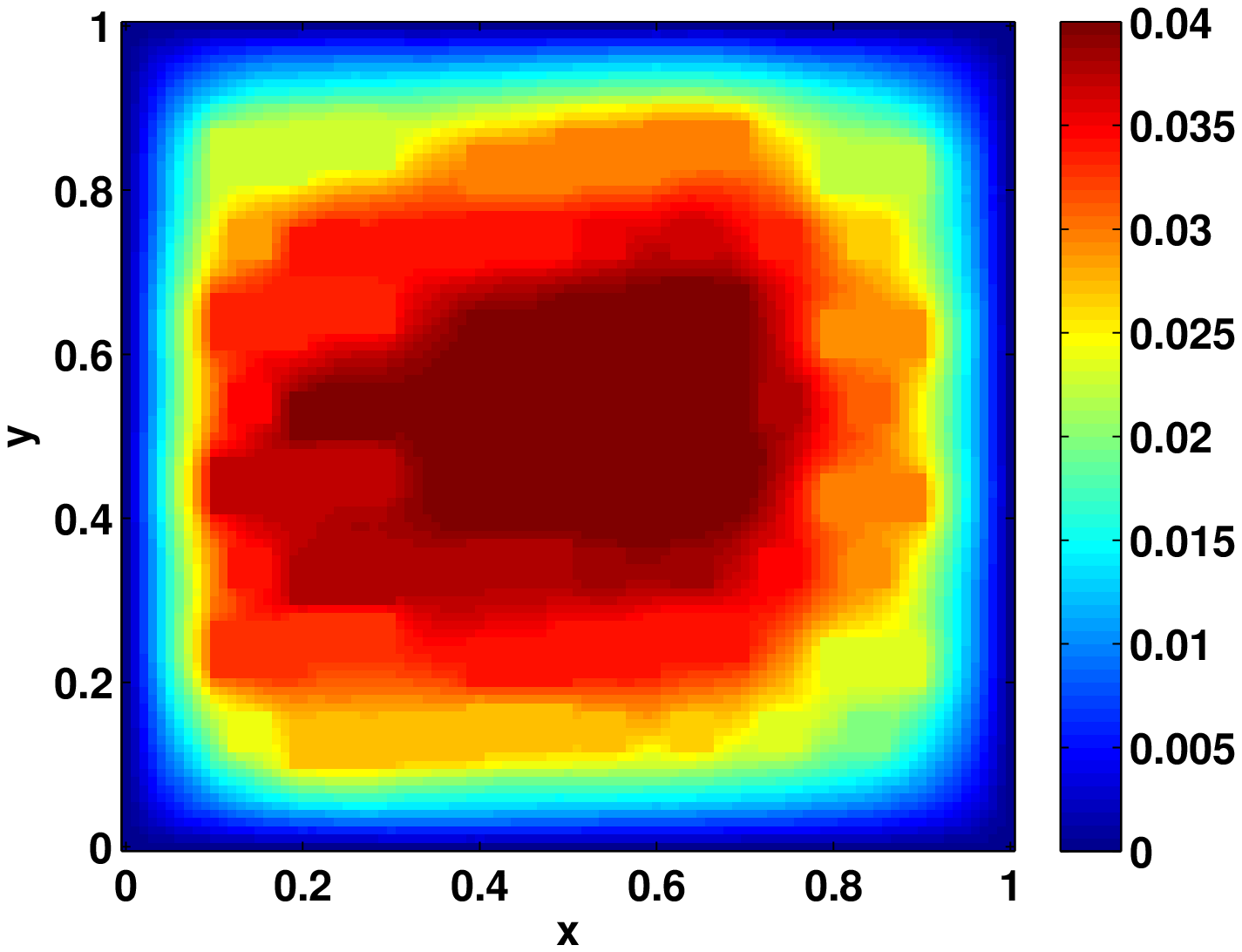}\label{Sol_High_Final_Ref}}
      \subfigure[Coarse-DMD -Approximate solution ($N_r=6$)]{\includegraphics[width=0.32\textwidth]{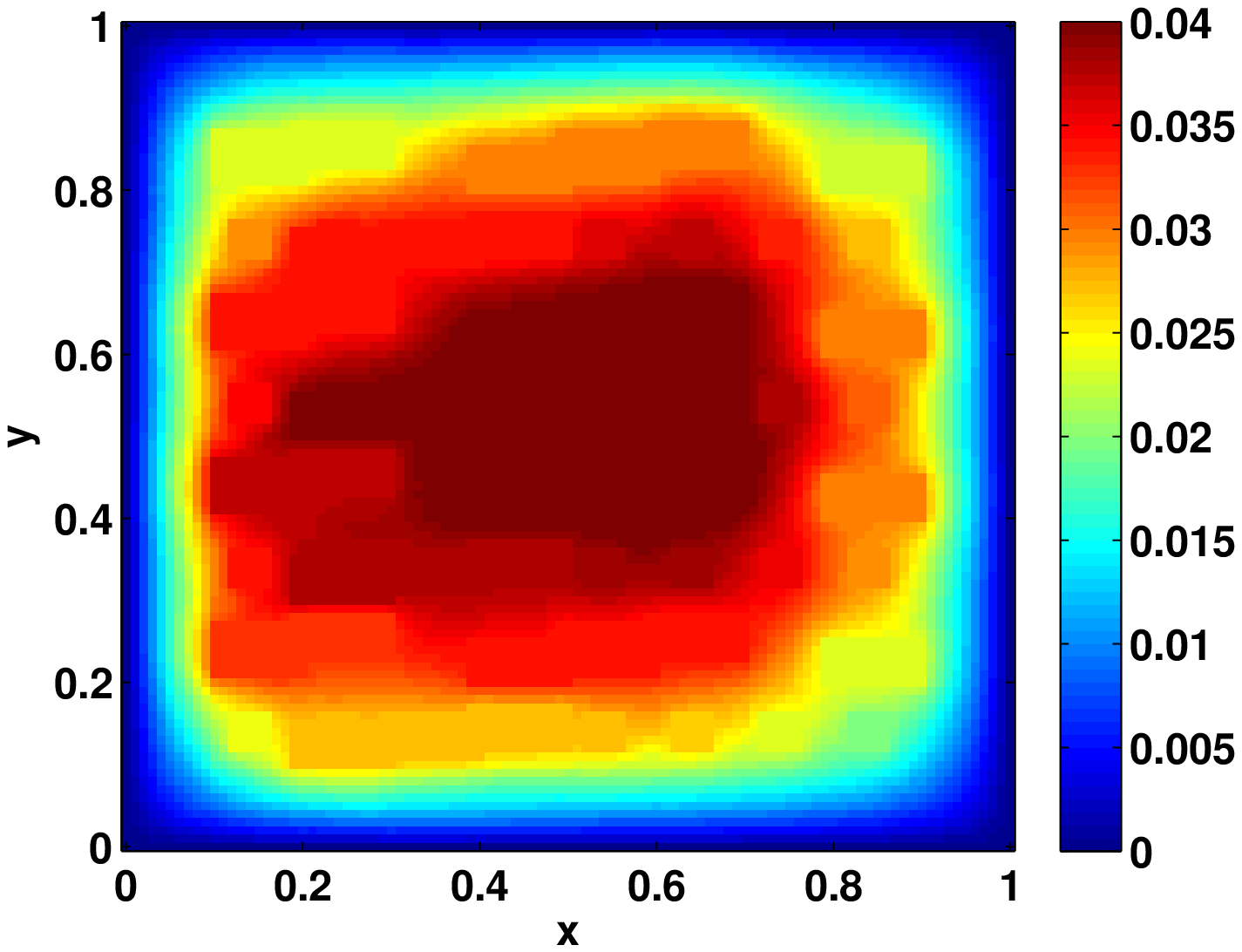}\label{Sol_High_Final_DMD}}
      \subfigure[Coarse-POD -Approximate solution ($N_r=6$)]{\includegraphics[width=0.32\textwidth]{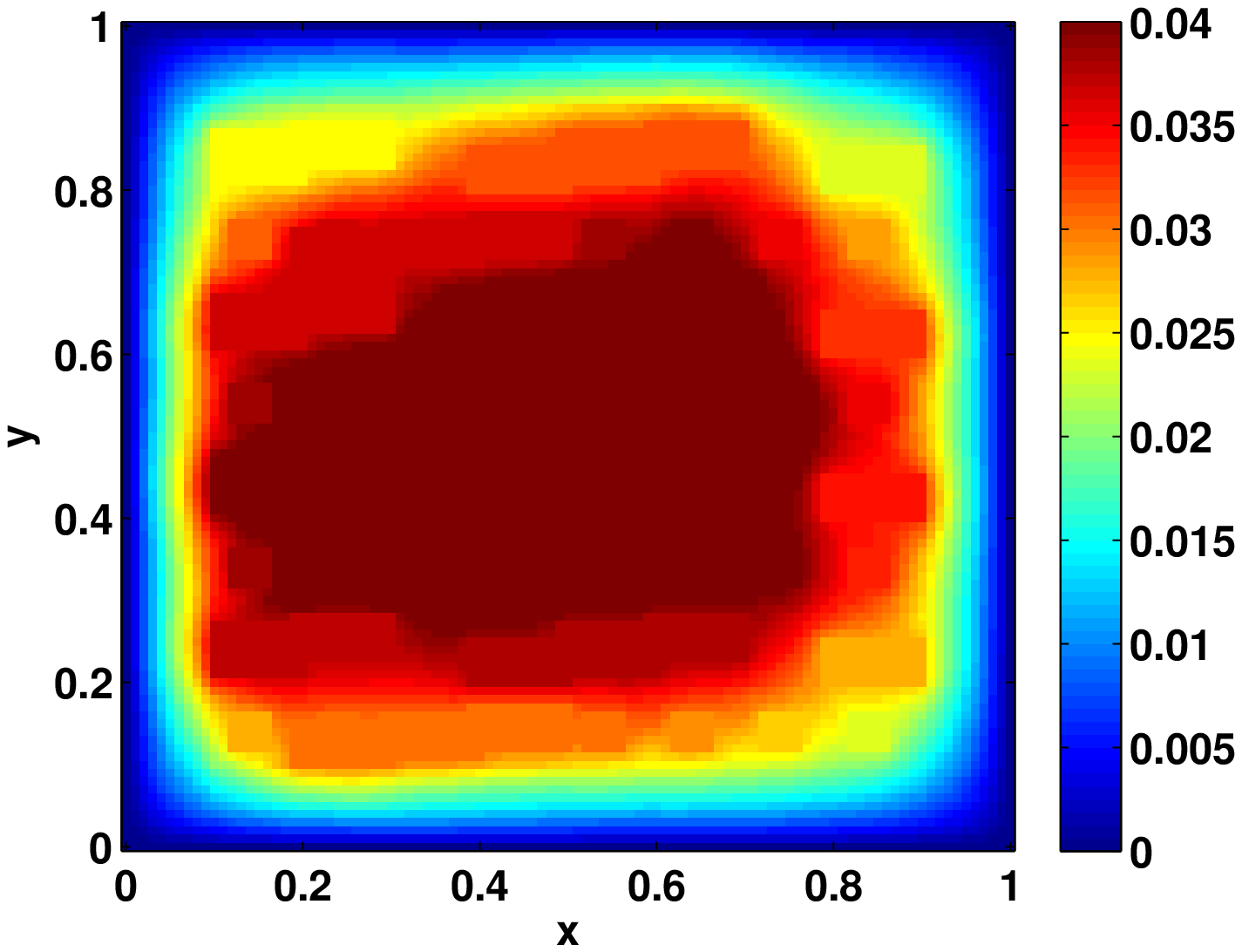}\label{Sol_High_Final_POD}}
  \end{center}
  \caption{Output fields at $t=1$ (Case I: high-conductivity channels). Comparison between reference solution of the fine scale problem with those obtained from the hybrid DMD- and POD-coarse multiscale approaches.}
  \label{Sol_High_Final}
\end{figure}

\begin{figure}[htb]
  \begin{center}
      \subfigure[Fine -Reference solution ($N_f=10201$)]{\includegraphics[width=0.32\textwidth]{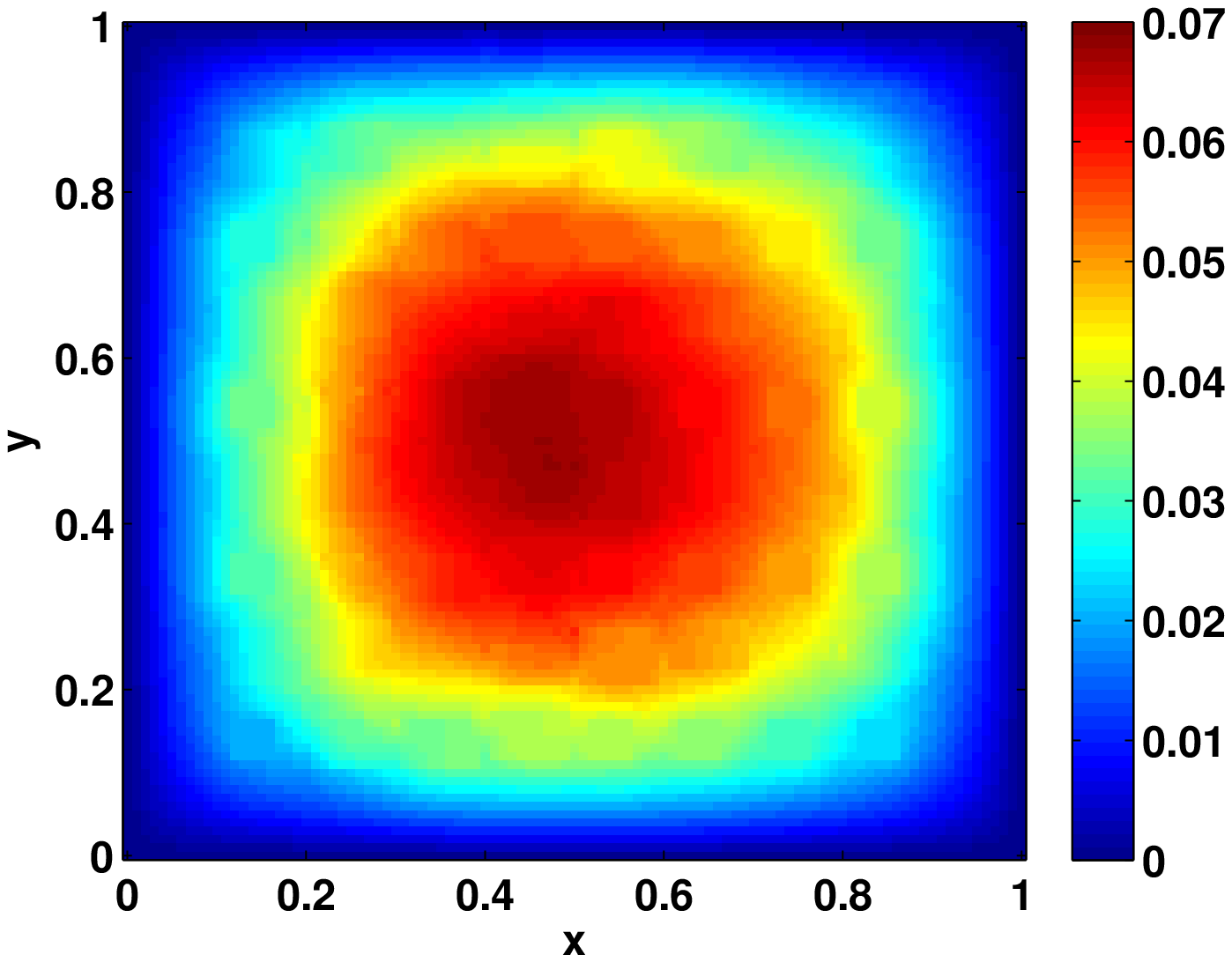}\label{Sol_Low_Final_Ref}}
      \subfigure[Coarse-DMD -Approximate solution ($N_r=6$)]{\includegraphics[width=0.32\textwidth]{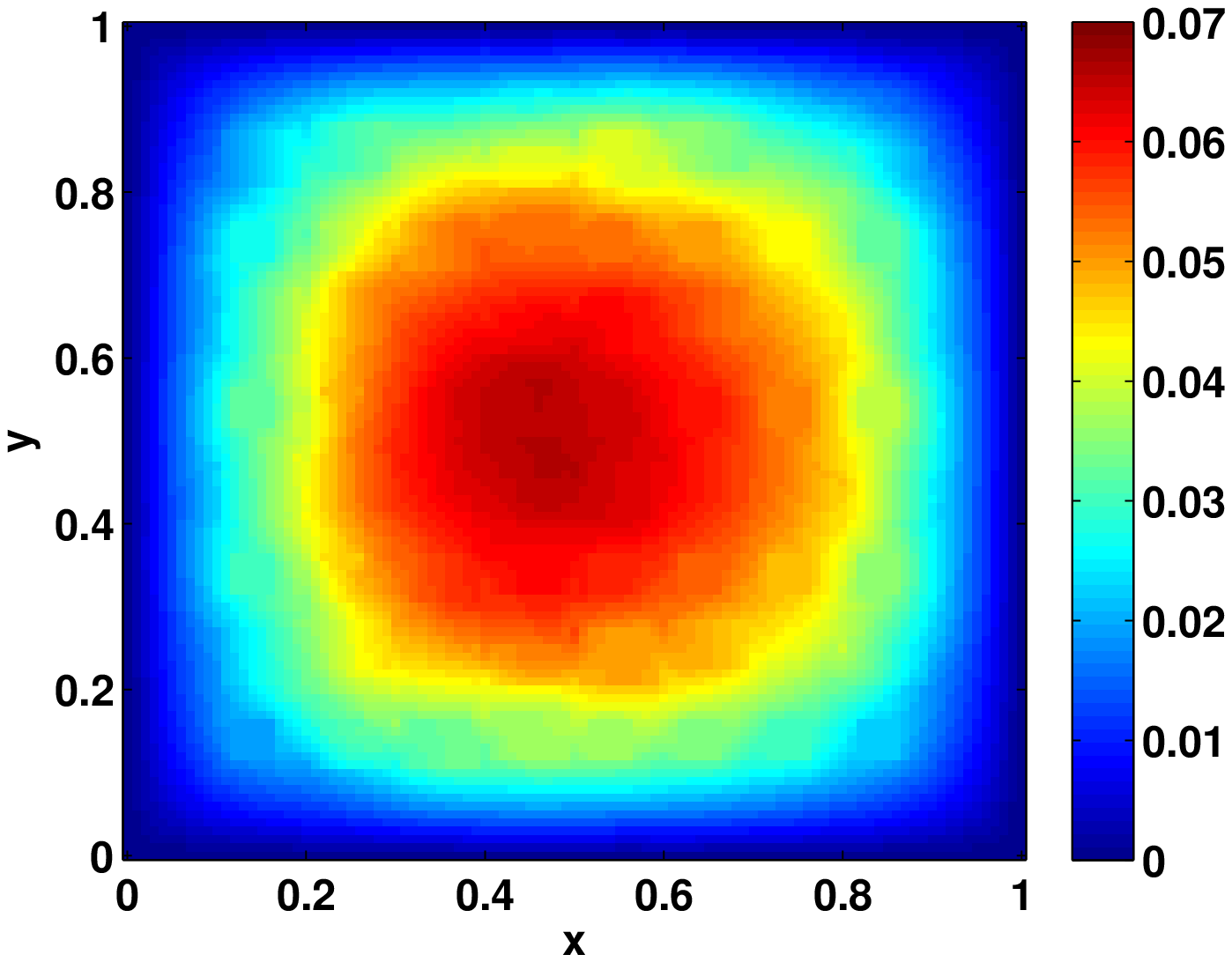}\label{Sol_Low_Final_DMD}}
      \subfigure[Coarse-POD -Approximate solution ($N_r=6$)]{\includegraphics[width=0.32\textwidth]{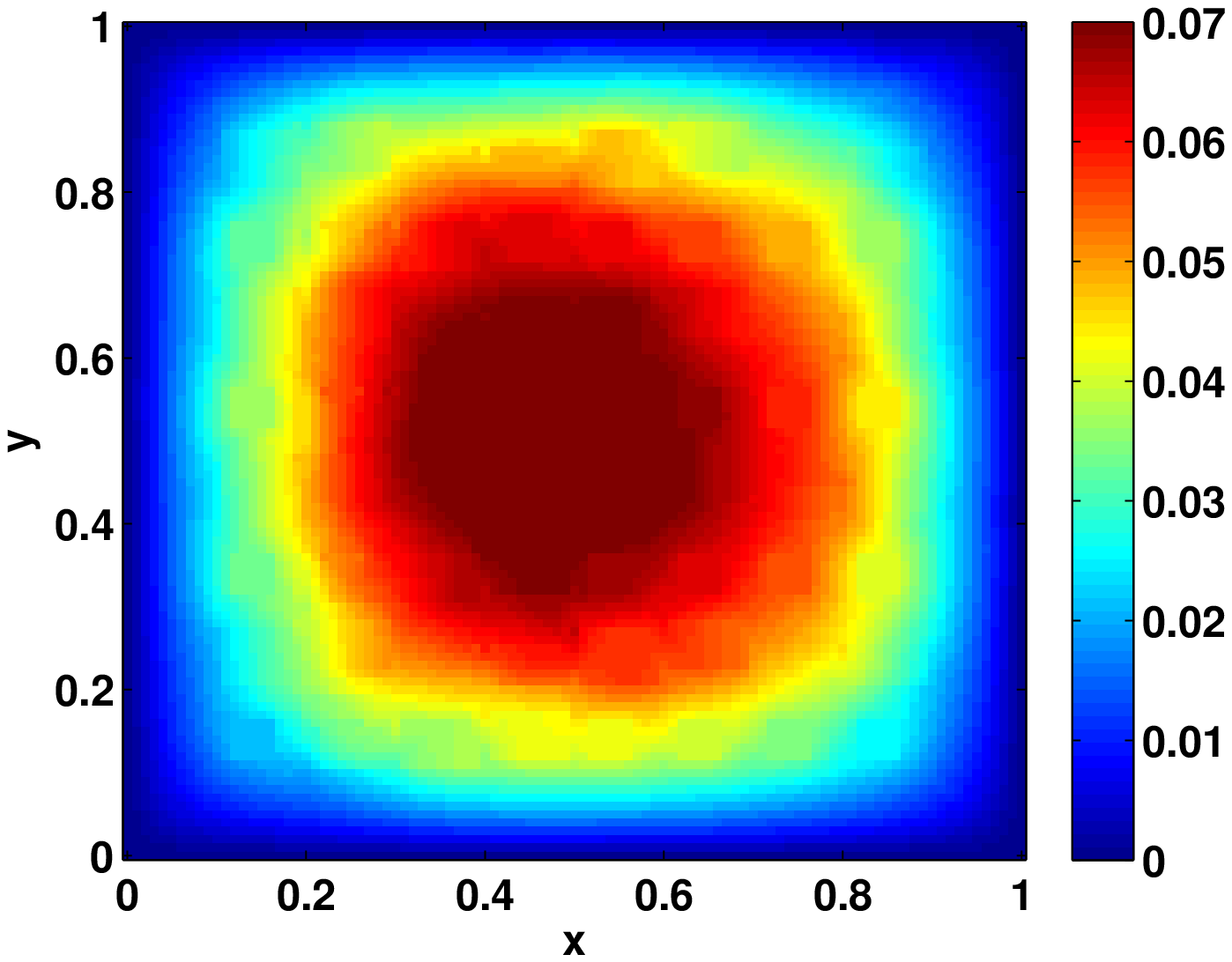}\label{Sol_Low_Final_POD}}
  \end{center}
  \caption{Output fields at $t=1$ (Case II: low-conductivity layers). Comparison between reference solution of the fine scale problem with those obtained from the hybrid DMD- and POD-coarse multiscale approaches.}
  \label{Sol_Low_Final}
\end{figure}

Figs. \ref{Sol_High_Final} and \ref{Sol_Low_Final} depict the output fields at $t=1$ obtained from the reference fine scale solution and those obtained from the local-global approach for both high and low conductivity cases. We observe good agreement between the reference solution and that obtained from the hybrid DMD-coarse multiscale approach. However, a discrepancy can be seen when comparing the reference solution with that obtained from the hybrid POD-coarse multiscale approach. These observations show the capability of DMD modes, computed from the first few snapshots, to capture the relevant flow dynamics and forecast the flow field with good accuracy for long time periods.

\begin{figure}[htb]
  \begin{center}
      \includegraphics[width=0.65\textwidth]{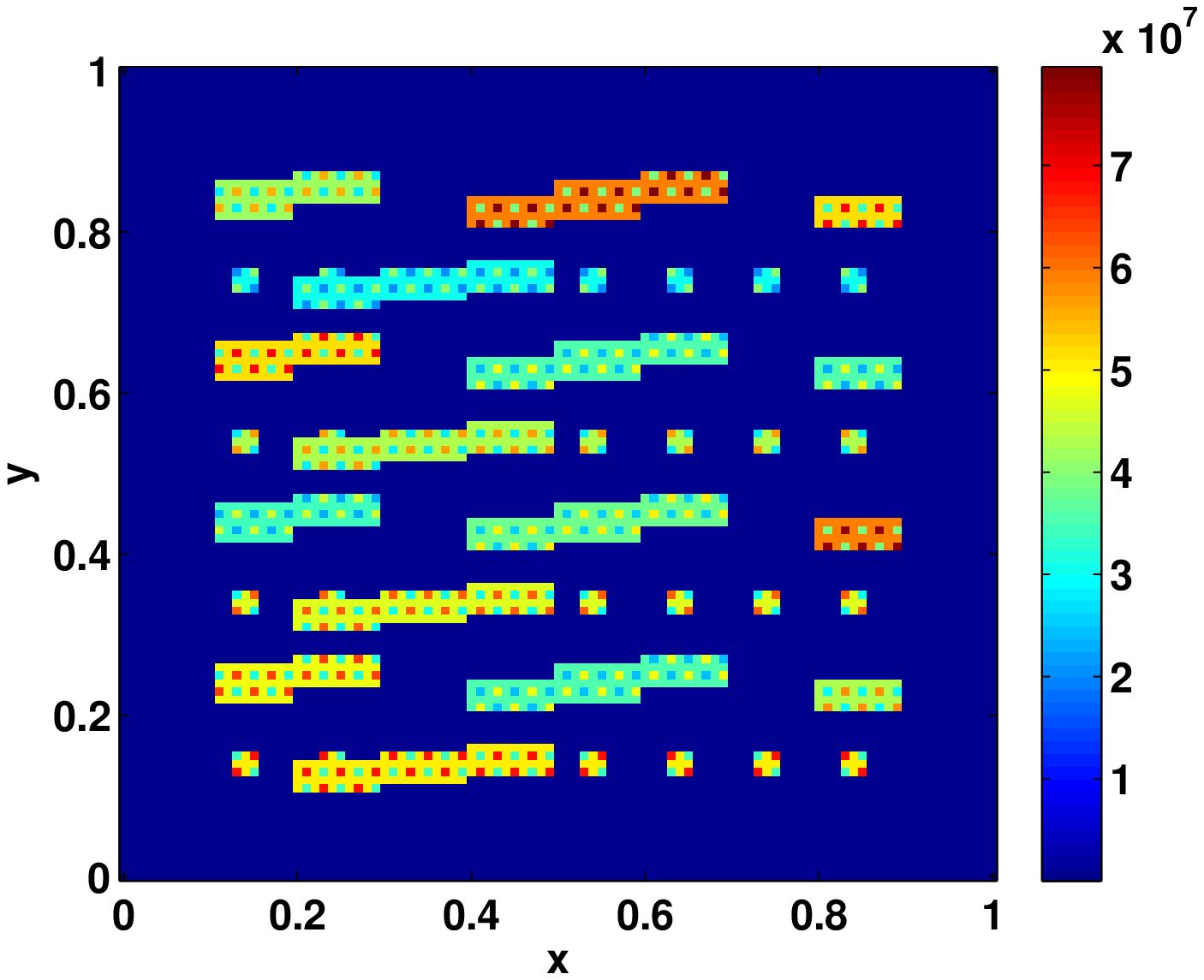}
  \end{center}
  \caption{Permeability field.}
  \label{Per_High_S}
\end{figure}

\begin{figure}[htb]
  \begin{center}
      \includegraphics[width=0.65\textwidth]{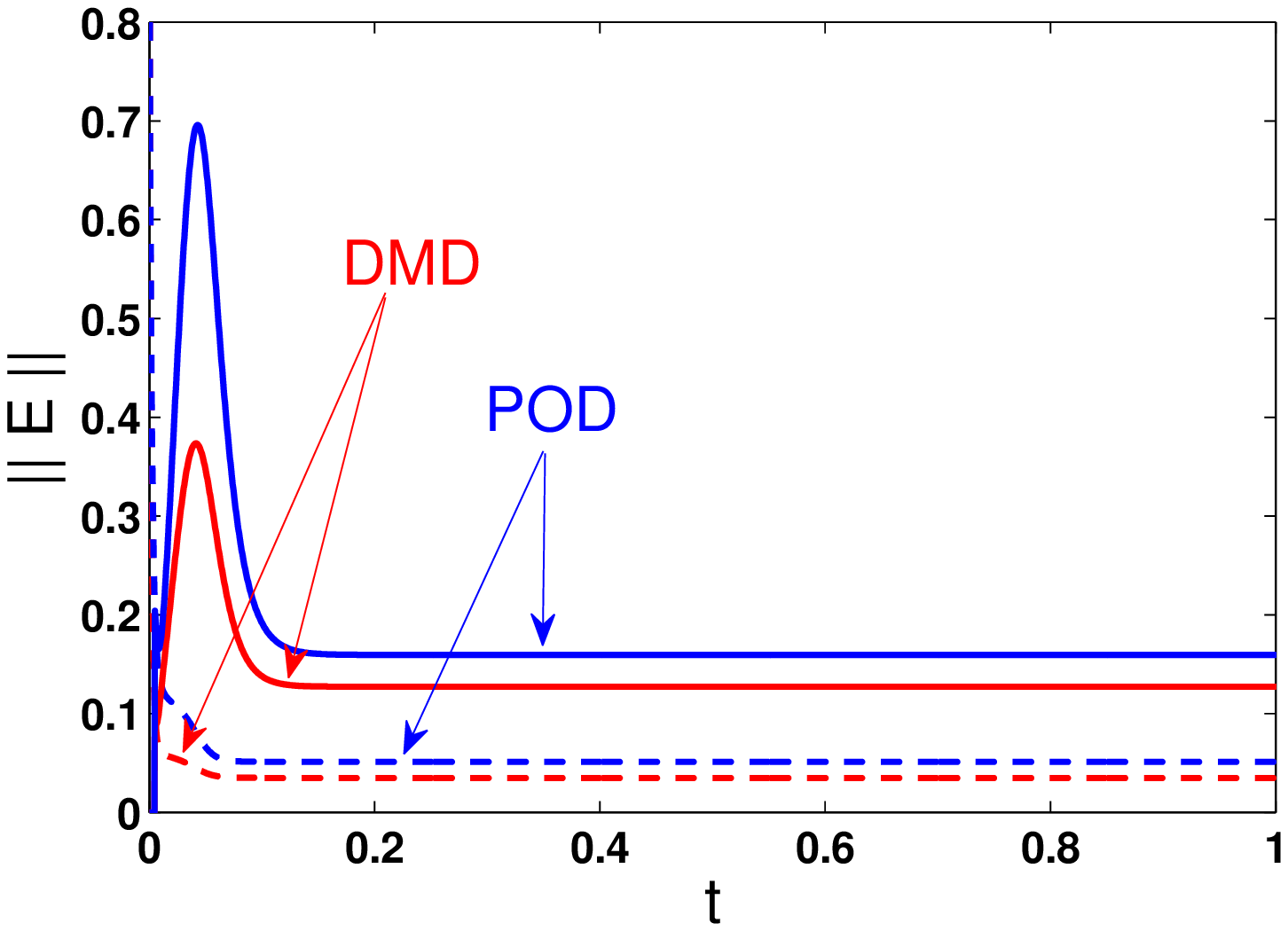}
  \end{center}
  \caption{Variations of the Galerkin projection error (solid line) and L$_2$ projection error (dashed line) with time for the perturbed permeability field shown in Fig. \ref{Per_High_S}. Results are obtained using POD and DMD modes.}
  \label{Error_HighCond_S}
\end{figure}

To investigate further the suitability of POD and DMD modes to model flows in varying and highly heterogeneous porous media, we consider Case I; that is, the permeability field shown in Fig. \ref{Per_High} and multiply its coefficient by a smooth positive spatial function to obtain
\begin{equation}\label{perm_smooth}
\kappa_s(x;y;\epsilon;f)=\kappa(x;y)\times(1+\epsilon+\sin(2\pi f x)\sin(2\pi f y) ),
\end{equation}
where $\epsilon =2$ and $f=25$. The resulting permeability field is depicted in Fig. \ref{Per_High_S}. This analysis is motivated by several applications which require solving the forward problem for varying permeability fields; for instance, when the permeability field is subject to uncertainty or multi-phase flow where the permeability is modulated by coarse-grid mobility. To address this issue, we use POD and DMD modes generated for the permeability field shown in Fig. \ref{Per_High} and employ the Galerkin projection to obtain a reduced-order model which is used to predict the flow field resulting from the modified permeability field described by Eq. (\ref{perm_smooth}) shown in Fig. \ref{Per_High_S}. In Fig. \ref{Error_HighCond_S}, we plot the temporal variations of the Galerkin projection and L$_2$ projection errors obtained from the POD- and DMD-based representations. Low steady-state errors are observed. In particular, smaller values are obtained when implementing the dynamic mode decomposition procedure on the coarse-scale problem. This shows the robustness of the hybrid DMD-coarse multiscale approach with respect to moderate perturbation in the permeability field.

\section{Conclusion}
\label{conclusion}
In this work, we propose a local-global approach for model reduction of high-contrast and time-dependent parabolic problems that govern flows in highly-heterogeneous porous media. This approach combines the concepts of generalized multiscale finite element method (GMsFEM) and proper orthogonal decomposition (POD) and/or dynamic mode decomposition (DMD) techniques. We consider different high-contrast coefficients and present numerical results to investigate the capability of our proposed approach to accurately capture the behavior of resolved solutions. The hybrid DMD-GMsFEM technique shows great potential to reproduce the flow field with good accuracy while reducing significantly the size of the original problem. This is due to the systematic construction of accurate coarse spaces from GMsFEM along with DMD's ability to extract the dynamic information and especially the relevant modes that govern the long-time dynamics. This achievement opens the door for large-scale optimization applications where the stochasticity of the parameters as well as their uncertainty can be efficiently involved in large-scale and accurate simulations.  

\bibliography{RefDec2012}
\bibliographystyle{elsarticle-num}

\end{document}